\documentclass[12pt]{article}

\usepackage{graphics}

\usepackage{amsfonts,amsmath,amsthm,stmaryrd,enumerate,a4wide}

\newtheorem{open}{Open question}
\newtheorem{thrm}{Theorem}[section]
\newtheorem{clm}[thrm]{Claim}
\newtheorem{lmm}[thrm]{Lemma}
\newtheorem{prpstn}[thrm]{Proposition}
\newtheorem{crllr}[thrm]{Corollary}

\theoremstyle{definition}

\newtheorem{dfntn}[thrm]{Definition}
\newtheorem{rmrk}[thrm]{Remark}
\newtheorem{xmpl}[thrm]{Example}

\newcommand{\textiff}{i{f}f}

\newcommand{\carre}[2]{#1^{#2 \times #2}}
\newcommand{\defeq}{\mathrel{\mathop:}=}
\newcommand{\sur}{\mathbin{/}}

\newcommand{\N}{\mathbb{N}}
\newcommand{\Q}{\mathbb{Q}}
\newcommand{\C}{\mathbb{C}} 
\newcommand{\Z}{\mathbb{Z}} 
\newcommand{\R}{\mathbb{R}} 
\newcommand{\FreeM}{\mathbb{W}}
\newcommand{\Hamilton}{\mathcal{H}}

\newcommand{\qmark}{\bigcirc} 

\newcommand{\Sigmabar}{\Sigma \cup \bar \Sigma}

\newcommand{\FG}[1]{\mathop{\mathrm{FG}}(#1)}

\newcommand{\lgr}[1]{\left| #1 \right|}
\newcommand{\nocc}[2]{\lgr{#2}_{#1}}
\newcommand{\calx}{\mathcal{X}}
\newcommand{\calz}{\mathcal{Z}}

\newcommand{\caly}{\mathcal{Y}}

\newcommand{\calf}{\mathcal{F}}
\newcommand{\calp}{\mathcal{P}}
\newcommand{\calc}{\mathcal{C}}

  \newcommand{\bcalc}{{\calc}'}
\newcommand{\ruset}{recursive underlying set}

\newcommand{\bfx}{\mathbf{x}}
\newcommand{\bfy}{\mathbf{y}}

\newcommand{\tta}{\mathtt{a}}
\newcommand{\ttb}{\mathtt{b}}
\newcommand{\tte}{\mathtt{e}}
\newcommand{\ttd}{\mathtt{d}}

 \newcommand{\ttx}{\mathtt{x}}
 \newcommand{\tty}{\mathtt{y}}
 \newcommand{\ttz}{\mathtt{z}}
\newcommand{\ttI}{\mathtt{I}}
\newcommand{\ttQ}{\mathtt{Q}}
\newcommand{\ttT}{\mathtt{T}}

\newcommand{\seg}[2]{\llbracket #1, #2 \rrbracket}  
\newcommand{\GL}[2]{\mathrm{GL}(#1, #2)}

\newcommand{\smalltwotwo}[4]
{ \left[ \begin{smallmatrix} #1 & #2 \\ #3 & #4 \end{smallmatrix} \right] }

\newcommand{\twotwo}[4]
{ \begin{bmatrix} #1 & #2 \\ #3 & #4 \end{bmatrix}  }

\newcommand{\twoone}[2]
{ \begin{bmatrix} #1 \\ #2  \end{bmatrix}}

\newcommand{\smalltwoone}[2]
{ \left[ \begin{smallmatrix} #1 \\ #2  \end{smallmatrix} \right] }

\newcommand{\onetwo}[2]
{ \begin{bmatrix} #1 & #2  \end{bmatrix} }

\newcommand{\Tri}{\mathrm{Tri}}

  \newcommand{\transit}[3]{#1 \mathrel{\xrightarrow{\ {#2} \ }} #3}

\newcommand{\ze}{\mathtt{0}}
\newcommand{\on}{\mathtt{1}}
\newcommand{\tw}{\mathtt{2}}
\newcommand{\zeon}{\left\{ \ze, \on \right\}}
\newcommand{\zobed}{\left\{ \ze, \on,  \ttb, \tte, \ttd \right\}}
\newcommand{\zeonzeon}{\FreeM \times \FreeM}
\newcommand{\mv}{\varepsilon}

\newcommand{\pbfree}[1]{\textsc{Free}$\big[ #1 \big]$}
\newcommand{\pbkfree}[2]{\textsc{Free}$(#1)\big[ #2 \big]$}
\newcommand{\pbtorsion}{\textsc{Matrix Torsion}}
\newcommand{\pbmortor}{\textsc{Morphism Torsion}}
\newcommand{\pbmpb}{\textsc{Matrix Power Boundedness}}

\newcommand{\pbkmatpow}[1]{\textup{MP}$(#1)$}
\newcommand{\pbaccess}{\textsc{Access}}
\newcommand{\pbkaccess}[1]{\pbaccess$(#1)$}

\newcommand{\pbratmem}[1]{\textsc{Accept}$\big[ #1 \big]$}
\newcommand{\pbsemmem}[1]{\textsc{Member}$\big[#1\big]$} 
\newcommand{\pbksemmem}[2]{\textsc{Member}$(#1)\big[#2\big]$} 
\newcommand{\pbsemfin}[1]{\textsc{Finite}$\big[#1\big]$}
\newcommand{\pbksemfin}[2]{\textsc{Finite}$(#1)\big[#2\big]$}
\newcommand{\pbbound}[1]{\textsc{Bounded}$\big[#1\big]$}
\newcommand{\pbkbound}[2]{\textsc{Bounded}$(#1)\big[#2\big]$}
\newcommand{\pbmort}[1]{\textsc{Mortal}$\big[#1\big]$}
\newcommand{\pbkmort}[2]{\textsc{Mortal}$(#1)\big[#2\big]$}

\newcommand{\pbGPCP}{\textup{GPCP}}
\newcommand{\pbkGPCP}[1]{\pbGPCP$(#1)$}
\newcommand{\pbPCP}{\textup{PCP}}
\newcommand{\pbkPCP}[1]{\pbPCP$(#1)$}
\newcommand{\pbMMPCP}{\textup{MMPCP}}
\newcommand{\pbkMMPCP}[1]{\pbMMPCP$(#1)$}

\newcommand{\rmD}{\mathrm{D}}
 \newcommand{\rmF}{\mathrm{F}}

\newcommand{\mvg}{1_G}

\begin{document}

\title{On the decidability of semigroup freeness}
\author{Julien Cassaigne and Francois Nicolas}

\maketitle


\begin{abstract}
This paper deals with the decidability of semigroup freeness.
More precisely, the freeness problem over a semigroup $S$ is defined as: 
given a finite subset $X \subseteq S$, 
decide whether each element of $S$ has at most one factorization over $X$.
To date, the decidabilities of the following two freeness problems have been closely examined. 
In 1953, Sardinas and Patterson proposed a now famous algorithm for the freeness problem over the free monoids.
In 1991, Klarner, Birget and Satterfield proved the undecidability of the freeness problem over three-by-three integer matrices.
Both results led to the publication of many subsequent papers.  

The aim of the present paper is $(i)$ to present general results about freeness problems,
$(ii)$ to study the decidability of freeness problems over various particular semigroups
(special attention is devoted to multiplicative matrix semigroups), 
and $(iii)$ to propose precise, challenging  open questions in order to promote the study of the topic.
\end{abstract} 

\maketitle 

\sloppy

\section{Introduction} \label{sec:intro}

We  first introduce  basic notation and definitions;
the organization of the paper is more precisely described in Section~\ref{sec:contrib}.

As usual, $\N$, $\Z$, $\Q$, $\R$, and $\C$ denote 
the semiring of naturals, 
the ring of integers, 
the field of rational numbers,
the field of real numbers, and 
the field of complex numbers, 
respectively.
For all  $m$, $n \in \Z$, 
$\seg{m}{n}$ denotes the set of all $k \in \N$ such that $m \le k \le n$.
Unless otherwise stated, 
the additive and multiplicative identity elements of any semiring are simply denoted $0$ and $1$, respectively.
The letter $O$ denotes any matrix  whose entries are all $0$. 

A \emph{word} is a finite sequence of symbols called its letters.
The \emph{empty word} is denoted  $\mv$.
For every word $w$,  the \emph{length} of $w$ is denoted $\lgr{w}$;
for every symbol $a$, 
$\nocc{a}{w}$ denotes the number of occurrences of $a$ in $w$.
An \emph{alphabet} is a (finite or infinite) set of symbols.
The canonical alphabet is the binary alphabet $\zeon$.

\subsection{Free semigroups and codes}

\subsubsection{Definitions}

A \emph{semigroup} is a set equipped with an associative binary operation.
Unless otherwise stated, semigroup operations are denoted multiplicatively.

\begin{dfntn}[Code]
Let $S$ be a semigroup and let $X$ be a subset of $S$.
We say that $X$ is a \emph{code} if the property
$$
x_1 x_2 \dotsm x_m  = y_1 y_2  \dotsm y_n
\iff 
(x_1, x_2, \dotsc, x_m) = (y_1, y_2, \dotsc, y_n)
$$ 
holds 
for any integers $m$, $n \ge 1$ and any elements  $x_1$, $x_2$, \ldots, $x_m$, $y_1$, $y_2$, \ldots, $y_n \in X$.
\end{dfntn}

Note that $(x_1, x_2, \dotsc, x_m) = (y_1, y_2, \dotsc, y_n)$ means that both $m = n$ and $x_i = y_i$ for every $i \in \seg{1}{m}$.
Informally,
a set is not a code \textiff{} its elements satisfy a non-trivial equation.
Or, in other words, 
a subset $X$ of a semigroup $S$ is a code \textiff{} no element of $S$ has more than one factorization over $X$.

For every semigroup $S$ and every subset $X \subseteq S$, $X^+$ denotes the closure of $X$ under the semigroup operation:  
$X^+$ is the subsemigroup of $S$ generated by $X$,
and as such, it is equipped with the semigroup operation induced by the operation of $S$.

\begin{dfntn}[Free semigroup] 
A semigroup $S$ is called \emph{free} if there exists a code $X \subseteq S$ such that $S = X^+$. 
\end{dfntn} 

In other words, 
a semigroup is free \textiff{} it is generated by a code.

A semigroup with an identity element is called a \emph{monoid}.
Many semigroups mentioned in the paper are monoids.
For every monoid $M$ and every subset $X \subseteq M$, $X^\star$ denotes the set $X^+$ 
augmented with the identity element of $M$.
A monoid $M$ is called \emph{free}  if there exists a code $X \subseteq M$ such that $M = X^\star$.

\begin{rmrk} 
No monoid is a free semigroup.
\end{rmrk}

\subsubsection{Illustration}

Let $\Sigma$ be an alphabet.
The set of all words over $\Sigma$ is a free monoid under concatenation
with 
$\mv$ as identity element 
and 
$\Sigma$ as generating code.
In accordance with our notation, this monoid is denoted as usual $\Sigma^\star$. 
In the same way, the set of all non-empty words over $\Sigma$ equals $\Sigma^+$ and is a free semigroup.
Both examples of free monoid and free semigroup are canonical 
(see Section~\ref{sec:isomorphisme}).

A subset of $\Sigma^\star$ is called a \emph{language} over $\Sigma$.
In the context of combinatorics on words, 
the term ``code'' was originally introduced to denote those languages that are codes under concatenation.
This particular topic has been widely studied \cite{BerstelPR09}.
A \emph{prefix code} over $\Sigma$ is a subset $X \subseteq \Sigma^+$ such that for every  $x \in X$ and every $s \in \Sigma^+$, $xs \notin X$.
It is clear that every prefix code is a code under concatenation.

\begin{xmpl}
Consider the semigroup $\FreeM \defeq \zeon^\star$.
The three subsets $\left\{ \ze\ze, \ze \on, \on \ze, \on \on \right\}$,
$\left\{ \ze \on, \ze \on \on, \on\on \right\}$ 
and $\left\{ \ze^n \on : n \in \N \right\}$ of $\FreeM$ are codes under concatenation,
but $\{ \ze \on, \on \ze , \ze \}$ is not:  
$\ze (\on \ze) = (\ze \on )\ze$.
\end{xmpl}

For any two semigroups $S_1$ and $S_2$, 
define the \emph{direct product} of $S_1$ and $S_2$ as 
 the Cartesian product $S_1 \times S_2$ equipped with  the componentwise semigroup operation derived from the operations of $S_1$ and $S_2$: 
for any two elements $(x_1, x_2)$ and  $(y_1, y_2)$ of $S_1 \times S_2$,
the product $(x_1, x_2)(y_1, y_2)$ is defined as $(x_1y_1, x_2 y_2)$.

\begin{xmpl} 
Consider the semigroup $\zeonzeon$. 
Both subsets $\{ (\ze, \on), (\on, \ze) \}$ and  $\left\{  (\ze, \ze), (\on, \ze\on), (\ze\on, \on\ze) \right\}$ of $\zeonzeon$ are codes under componentwise concatenation but
 $\{ (\ze, \ze), (\on, \on \ze \on  ), (\ze \on, \ze \on) \}$ is not: 
$
(\ze, \ze)(\on, \on \ze \on  )(\ze \on, \ze  \on)  
= (\ze \on, \ze  \on) (\ze, \ze)(\on, \on \ze \on  )
$.
\end{xmpl}

Let $D$ be a semiring and let  $\carre{D}{d}$ denote the set of all $d$-by-$d$ matrices over $D$: 
$\carre{D}{d}$ is a semiring under the matrix operations induced by the operations of $D$, 
so in particular, 
$\carre{D}{d}$ is a multiplicative semigroup.

\begin{xmpl}
Consider the semigroup $\carre{\N}{2}$.
Let $k$ be an integer greater than $1$.
The subsets
$$\left\{ \twotwo{k}{i}{0}{1} : i \in \seg{0}{k - 1}  \right\}
$$
and
$$\left\{ \twotwo{1}{1}{0}{1}, \twotwo{1}{0}{1}{1} \right\}
$$
of $\carre{\N}{2}$ are codes under matrix multiplication  \cite{CassaigneHK99} 
but 
$$\left\{  \twotwo{1}{0}{0}{2}, \twotwo{1}{3}{0}{1}\right\}
$$
 is not: 
$$
 \twotwo{1}{3}{0}{1} \twotwo{1}{0}{0}{2} 
 = \twotwo{1}{0}{0}{2} 
 \twotwo{1}{3}{0}{1} \twotwo{1}{3}{0}{1} \, .
$$
\end{xmpl}

\subsection{Freeness problems}

Our aim is to study the decidability of freeness problems over various semigroups:

\begin{dfntn}
Let $S$ be a semigroup with a \ruset. 
The \emph{freeness problem over $S$}, denoted  \pbfree{S}, is: 
given a finite subset $X \subseteq S$, 
decide whether $X$ is a code.
For every integer $k \ge 1$, define \pbkfree{k}{S} as the following problem:
given a $k$-element subset $X \subseteq S$, 
decide whether $X$ is a code. 
\end{dfntn}

For every integer $k \ge 1$, \pbkfree{k}{S} is a restriction of \pbfree{S}.

\begin{rmrk}
Let $S$ be a semigroup with a \ruset. 
\pbfree{S}  should not be confused with the following problem,
which is not the concern of the paper:
 given a finite subset  $X \subseteq S$, decide whether $X^+$ is a free semigroup.
For any $a$, $b \in S$  such that $\{ a, b \}$ is a code,
$\{ a, b, ab \}$ is not a code but ${\{ a, b, ab \}}^+$ is a free semigroup.
In general, for every  subset of $X \subseteq S$, 
the semigroup $X^+$ is free \textiff{} there exists a code $Y \subseteq X$ such that $X \subseteq  Y^+$.
\end{rmrk}

Let us now present some relevant examples of freeness problems.

\begin{xmpl} \label{ex:sardine}
The decidability of \pbfree{\Sigma^\star} for any finite alphabet $\Sigma$ was proven by Sardinas and Patterson in 1953. 
Efficient polynomial-time algorithms were proposed afterwards \cite{BerstelPR09}.
\end{xmpl}

\begin{xmpl} \label{ex:zeon-d}
For any alphabet $\Sigma$ and any $x$, $y \in \Sigma^\star$ with $x \ne y$, the following three assertions are equivalent:
\begin{enumerate}
 \item $\{ x, y \}$  is not a code, 
 \item $xy = yx$, and 
 \item there exist $s \in \Sigma^\star$ and  $p$, $q \in \N$ such that $x = s^p$ and $y = s^q$ \cite{Lothaire1,Lothaire2,BerstelPR09}.
\end{enumerate}
More generally,  
let $\Sigma_1$, $\Sigma_2$, \ldots, $\Sigma_d$ be $d$ alphabets, 
and  
let $\bfx =  (x_1, x_2, \dotsc, x_d)$ and $\bfy = (y_1, y_2, \dotsc, y_d)$ be two elements of $\Sigma_1^\star \times \Sigma_2^\star \times \dotsb \times \Sigma_d^\star$.
The $2$-element set 
$\left\{  \bfx, \bfy \right\}$
is not a code \textiff{} $x_i y_i = y_i x_i$ for every $i \in \seg{1}{d}$.
Hence, if $\Sigma_i$ is finite for every $ i \in \seg{1}{d}$ then \pbkfree{2}{\Sigma_1^\star \times \Sigma_2^\star \times \dotsb \times \Sigma_d^\star} is decidable. 
In Section~\ref{sec:three-three}, 
we prove that \pbfree{\zeonzeon} is undecidable.
\end{xmpl}

\begin{xmpl}  \label{ex:mat-carre-free}
For each  integer $d \ge 1$,  
\pbkfree{1}{\carre{\Q}{d}} is decidable in polynomial time \cite{MandelS77} (see also Section~\ref{sec:one-mat}).
However, 
Klarner, Birget, and Satterfield proved in 1991 that \pbfree{\carre{\N}{3}}  is undecidable.
More precisely, \pbkfree{k}{\carre{\N}{3}} is decidable for at most finitely many integers $k \ge 1$ \cite{CassaigneHK99,Claus07} (see also Section~\ref{sec:three-three}). 
\end{xmpl}

\subsection{Contribution} \label{sec:contrib}

The paper is divided into eight sections.

\paragraph{Section~\ref{sec:intro}.}

In the remainder of this section, 
we first state some useful, basic facts about semigroup morphisms (Section~\ref{sec:morphism}).
Then, 
we present a list of previously studied problems related to the combinatorics of semigroups 
(Section~\ref{sec:def-other-pb}):
all along the paper, we compare their properties to those of freeness problems.

\paragraph{Section~\ref{sec:one-mat}.}

A square matrix $M$ is called torsion if there exist two integers $p$, $q \ge 1$ such that $M^p = M^{p + q}$;
equivalently, $M$ is torsion \textiff{} the singleton $\{ M \}$ is not a code under matrix multiplication.
We examine several questions related with matrix torsion.

\paragraph{Section~\ref{sec:balanced-eq}.}  

We first prove that for any semigroup $S$ and any subset $X \subseteq S$ with cardinality greater than $1$, 
$X$ is not a code \textiff{} the elements of $X$ satisfy a non-trivial \emph{balanced} equation.
We then explore the consequences of the latter statement.  
The most interesting of them is that,
for every integer $d \ge 1$, 
\pbfree{\carre{\Q}{d}} reduces to \pbfree{\carre{\Z}{d}}. 

\paragraph{Section~\ref{sec:semigroup-group}.}

We show that \pbfree{\GL{2}{\Z}} is decidable, 
and  that for every finite alphabet $\Sigma$, 
\pbfree{\FG{\Sigma}} is decidable in polynomial time, 
where $\FG{\Sigma}$ denotes the free group over $\Sigma$.
The latter result generalizes Example~\ref{ex:sardine}.
Both proofs rely on automata theory.

\paragraph{Section~\ref{sec:nbGene}.} 

We first show that the following seemingly obvious statement is wrong: 
for any semigroup $S$ with a \ruset{} and any integer $k \ge 1$, 
the decidability of \pbkfree{k + 1}{S}  implies the decidability of \pbkfree{k}{S}.
We then prove that,  
for any semigroup $S$ with a computable operation, 
either \pbkfree{k}{S}  is decidable for every integer $k \ge 2$, 
or 
\pbkfree{k}{S} is undecidable for infinitely many integers $k \ge 2$. 

\paragraph{Section~\ref{sec:dec-two-two}.}

The decidability of \pbfree{\carre{\N}{2}} is a very  exciting but difficult open question \cite{GawrychowskiGK10,BlondelCK04,CassaigneHK99,Klarner91}.
We propose new ideas to tackle the problem.

\paragraph{Section~\ref{sec:three-three}.} 

We prove that both \pbkfree{13 + h}{\zeonzeon}
and 
\pbkfree{13 + h}{\carre{\N}{3}} are undecidable for every $h \in \N$.
The undecidabilities of 
\pbkfree{13}{\zeonzeon}
and 
\pbkfree{13}{\carre{\N}{3}} were previously unknown.

\paragraph{Section~\ref{sec:deux-matrices}.}

We complete the picture of undecidability for freeness problems over matrix semigroups: 
we prove that 
\pbkfree{7 + h}{\carre{\N}{6}}, 
\pbkfree{5 + h}{\carre{\N}{9}}, 
\pbkfree{4 + h}{\carre{\N}{12}},
\pbkfree{3 + h}{\carre{\N}{18}}, and 
\pbkfree{2 + h}{\carre{\N}{36}} 
are undecidable for every $h \in \N$.

\paragraph{Open questions.} 

Relevant open questions are stated all along the paper.

\subsection{Semigroup morphisms} \label{sec:morphism}

\subsubsection{Definition}

Let $S$ and $S'$ be two semigroups.
A function $\sigma\colon S \to S'$ is called a \emph{morphism} if for all $x$, $y \in S$, $\sigma(xy) = \sigma(x) \sigma(y)$.
Note that even if both $S$ and $S'$ are monoids,
 a morphism from $S$ to $S'$ does not necessarily map the identity element of $S$ to the identity element of $S'$:
 ``morphism'' always mean ``semigroup morphism'' but not necessarily  ``monoid morphism''.
The following two  claims are  explicitly or implicitly used many times throughout the paper.

\begin{clm}[Universal property] \label{claim:def-morphism}
Let $\Sigma$ be an alphabet and let $S$ be a semigroup.
For any  function $s\colon \Sigma \to S$,
there exists exactly one morphism $\sigma\colon \Sigma^+ \to S$ such that $\sigma(a) = s(a)$ for every $a \in \Sigma$.
\end{clm}

\begin{clm} \label{clm:pre-image-code}
Let $S$ and $S'$ be two semigroups, 
let $\sigma\colon S \to S'$ be a morphism, and 
let $X$ be a subset of $S$.
The following two assertions are equivalent: 
\begin{enumerate}
 \item  $\sigma$ is injective on $X$ and $\sigma(X)$ is a code.
 \item $\sigma$ is injective on $X^+$ and $X$ is a code.
\end{enumerate}
\end{clm}

\subsubsection{Freeness problems as morphism problems}

Let $S$ be a semigroup with a \ruset{} and let  $\Sigma$ be a  finite alphabet.
Although the set of all functions from $\Sigma^+$ to $S$ has the power of the continuum  whenever $S$ is non-trivial,
the restriction of $\sigma$ to $\Sigma$ provides a finite encoding of $\sigma$ for any morphism $\sigma\colon \Sigma^+ \to S$. 
From now on such encodings are considered as canonical.
Hence, \pbfree{S} can be restated as follows: 
given a finite alphabet $\Sigma$ and a morphism $\sigma\colon \Sigma^+ \to S$, 
decide whether $\sigma$ is injective. 
In the same way, for every integer $k \ge 1$, an alternative formulation of  \pbkfree{k}{S} is: 
given an alphabet $\Sigma$ with cardinality $k$ and a morphism $\sigma\colon \Sigma^+ \to S$, 
decide whether $\sigma$ is injective.

\subsubsection{The free semigroup and the free monoid structures} \label{sec:isomorphisme}

A bijective morphism is called an \emph{isomorphism}.
The inverse function of any isomorphism  is also an isomorphism.
A semigroup  $S$ is free \textiff{}
for some alphabet $\Sigma$, 
there exists an isomorphism from  $\Sigma^+$ onto $S$.
A monoid $M$ is free \textiff{}
for some alphabet $\Sigma$, 
there exists an isomorphism from  $\Sigma^\star$ onto $M$.
Given a monoid $M$ and an alphabet $\Sigma$, 
every morphism from $M$ to $\Sigma^\star$ maps the identity element of $M$ to the empty word.
Since  $\FreeM$ contains infinite codes, \emph{e.g.}, the prefix code $\left\{  \ze^n \on  : n \in \N \right\}$, we may state:

\begin{clm} \label{claim:inject-zeon}
For any finite or countable alphabet $\Sigma$, 
there exists an injective morphism from $\Sigma^\star$ to $\FreeM$. 
\end{clm}

\subsection{Other decision problems} \label{sec:def-other-pb}

The decision problems that are stated in this section are related to the combinatorics of  semigroups.
Although they do not play any crucial role in the paper, 
it is interesting to compare their properties with the ones of the freeness problems.

\subsubsection{Semigroup membership and semigroup finiteness}

Let $S$ be a semigroup with a \ruset.
Define \pbsemmem{S} as the following problem: 
given a finite  subset $X \subseteq S$ and an element $a \in S$,
 decide whether $a \in X^+$; 
for every integer $k \ge 1$, 
\pbksemmem{k}{S} denotes the restriction of \pbsemmem{S} to those instances $(X, a)$ such that the cardinality of $X$ equals $k$.
Define \pbsemfin{S} as the following problem: 
given a finite subset $X \subseteq S$, decide whether $X^+$ is finite;
for every integer $k \ge 1$, 
\pbksemfin{k}{S} denotes the restriction of \pbsemfin{S} to input sets $X$ of cardinality $k$.

\subsubsection{Mortality \cite{Paterson70}}

Let $S$ be a semigroup.
A \emph{zero element} of $S$ is an element $z \in S$ such that $z s = sz = z$ for every $s \in S$.
No semigroup has more than one zero element.
For every semigroup $S$ with a \ruset{} and a zero element, 
let \pbmort{S} denote the following problem: 
given a finite subset $X \subseteq S$, decide whether the zero element of $S$ belongs to  $X^+$;
for every integer $k \ge 1$,  
\pbkmort{k}{S} denotes the restriction of \pbmort{S} to input sets $X$ of cardinality $k$.

\subsubsection{Boundedness \cite{BlondelT00}}

Let $d$ be a positive integer.
A subset $X \subseteq \carre{\C}{d}$ is called \emph{bounded} if there exists a non-negative constant $b$ such that the modulus of any entry of any matrix in $X$ is at most $b$.
Let $S$ be a recursive subset of $\carre{\Q}{d}$.
Let \pbbound{S} denote 
 the following problem: 
 given a finite subset $X \subseteq S$,
decide whether $X^+$ is bounded;
for every integer $k \ge 1$, 
\pbkbound{k}{S} denotes the restriction of \pbbound{S} to input sets $X$ of cardinality $k$.

\subsubsection{Generalized Post correspondence problem \cite{EhrenfeuchtKR82}}

Let \pbGPCP{} denote the following problem: 
given a finite alphabet $\Sigma$,
two morphisms $\sigma$, $\tau\colon \Sigma^\star \to \FreeM$,
and $s$, $s'$, $t$, $t' \in \FreeM$, 
decide whether there exists $w \in \Sigma^\star$ such that $s \sigma(w) s' = t \tau(w) t'$;
for every integer $k \ge 1$, \pbkGPCP{k} denotes the restriction of \pbGPCP{} to those instances $(\Sigma, \sigma, \tau, s, s', t, t')$ such that the cardinality of $\Sigma$ equals $k$.

\section{The case of a single generator} \label{sec:one-mat}

\begin{dfntn}
Let $S$ be a semigroup.
An element $s \in S$ is called \emph{torsion} if the following four equivalent conditions are met.
\begin{enumerate}
\item The singleton $\{ s \}$ is not a code.
\item There exist two  integers $p$ and $q$ with $0 < p  < q$ such that $s^p = s^q$.
\item The semigroup $\left\{ s, s^2, s^3, s^4, \dotsc \right\}$ has  finite cardinality.
\item The sequence $(s, s^2, s^3, s^4, \dotsc)$ is eventually periodic.
\end{enumerate}
\end{dfntn}

For any semigroup $S$ with a \ruset, \pbkfree{1}{S} is the complementary problem of  \pbksemfin{1}{S}.

\subsection{Matrix torsion over the complex numbers} 

The next theorem characterizes those complex square matrices that are torsion.
The proof uses the following basic fact from linear algebra:

\begin{lmm}[Theorem~3.3.6 in \cite{HornJ90}] \label{lem:eigen-mult}
Let  $M$ be a complex square matrix and
let $\lambda$ be an eigenvalue of $M$.
The multiplicity of $\lambda$ as a root of the minimal polynomial of $M$
equals
the maximum order of a Jordan block of $M$ corresponding to $\lambda$.
\end{lmm}

\begin{thrm} \label{caract-torsion-C}
Let $d$ be a positive integer and let $M \in \carre{\C}{d}$.
The following four assertions are equivalent.
\begin{enumerate}[$(i)$.]
\item \label{assert:M-torsion}
The matrix $M$ is torsion. 
\item \label{assert:pol-min}
There exist  $v \in \seg{0}{d}$ and a finite set $U$  of roots of unity such that the minimal polynomial of $M$ equals 
$$
\ttz^v \prod_{u \in U} (\ttz - u) \, .
$$  
\item \label{assert:jordan}
There exist  a diagonal matrix $D$ and a nilpotent matrix $N$ such that 
every eigenvalue of $D$ is a root of unity and 
$$
\twotwo{D}{O}{O}{N}
$$
is a Jordan normal form of $M$. 
\item \label{assert:Md-Mnd}
There exists an integer  $n \ge 2$ such that $M^d = M^{nd}$.
\end{enumerate}
\end{thrm} 

\begin{proof}
\begin{trivlist}
\item$(\ref{assert:M-torsion}) \implies (\ref{assert:pol-min})$.
Assume that assertion~$(\ref{assert:M-torsion})$ holds.
Then, there exist two integers $p$ and $q$ with $0 \le p < q$ such that $M^p = M^q$.
Let $\mu(\ttz)$ denote the minimal polynomial of $M$.
Since $M^q - M^p$ is a zero matrix,
$\mu(\ttz)$  divides 
 $\ttz^q - \ttz^p = \ttz^{q - p}(\ttz^p - 1)$.
Therefore, 
$\mu(\ttz)$ can be written in the form 
$\mu(\ttz) = \ttz^v \prod_{u \in U} (\ttz - u)$ 
with $v \in \seg{0}{q - p}$ 
and 
$U \subseteq \left\{ u \in \C : u^p = 1 \right\}$.
Moreover, the Cayley-Hamilton theorem implies that  $\mu(\ttz)$ divides the characteristic polynomial  of $M$ which is of degree $d$,
so $v$ is not greater than $d$.
We have thus shown assertion~$(\ref{assert:pol-min})$.
\item$(\ref{assert:pol-min}) \implies  (\ref{assert:jordan})$.
It follows from Lemma~\ref{lem:eigen-mult} that 
assertions $(\ref{assert:pol-min})$ and $(\ref{assert:jordan})$ are equivalent.

\item$(\ref{assert:jordan}) \implies (iv)$.
Assume that assertion~$(iii)$ holds.
Then, there exist a non-singular matrix $P$, a diagonal matrix $D$, and a nilpotent matrix $N$ such that 
every eigenvalue of $D$ is a root of unity and 
$$
M = P \twotwo{D}{O}{O}{N}P^{-1} \, .
$$
Let $m$ be a positive integer such that $\lambda^m = 1$ for every eigenvalue $\lambda$ of $D$.
Clearly,  
$D^m$ is an identity matrix, and thus $D^{(m + 1)d} = {(D^m)}^d D^d = D^d$.
Moreover,
$N^{(m + 1)d}$ and $N^{d}$ are equal to the same zero matrix, and thus
$$
M^{(m + 1)d} = P 
\twotwo{D^{(m + 1)d}}{O}{O}{N^{(m + 1)d}}
P^{-1} 
= 
 P 
\twotwo{D^d}{O}{O}{N^d}
P^{-1} 
= 
M^{d} \, .
$$
Hence, assertion~$(\ref{assert:Md-Mnd})$ holds with $n \defeq m + 1$.
\item$(\ref{assert:Md-Mnd}) \implies (\ref{assert:M-torsion})$.
Clearly,  
assertion~$(\ref{assert:Md-Mnd})$ implies assertion~$(\ref{assert:M-torsion})$.
 \qedhere
\end{trivlist}
\end{proof}

Let us now turn to matrices with rational entries.
The next proposition characterizes those two-by-two rational matrices that are torsion.

\begin{lmm} \label{lem:Euler-naif}
 Let $\phi$ denote Euler's totient function:  
for every integer $n \ge 1$, 
$\phi(n)$ equals the number of $k \in \seg{1}{n}$ such that $k$ and $n$ are coprime.
For every integer $n \ge 1$,
$\phi(n) = 2$ is equivalent to $n \in \{ 3, 4, 6 \}$.
\end{lmm}

\begin{proof}
Let $n$ be an integer greater than $6$.
Let $T$ denote the set of all $k \in \seg{1}{n}$ such that $k$ and $n$ are coprime.
Let $r$, $m \in \N$ be such that $n = 2^r m$ and $m$ is odd.
If $m \in \{ 1, 3 \}$ then $\{ 1, 5, n - 1 \}$ is $3$-element subset of $T$;
if $m \ge 5$ then $\{ 1, m - 2, n - 1 \}$ is a $3$-element subset of $T$.
Hence, $\phi(n)$  is greater than $2$ for every integer $n > 6$.
Besides, we have $\phi(1) = \phi(2) = 1$, $\phi(3)  = \phi(4) = \phi(6) = 2$, and $\phi(5) = 4$,  
so the lemma holds.
\end{proof}

Recall that for each integer $n \ge 1$, the degree of the $n^\text{th}$ cyclotomic polynomial, 
denoted $\Phi_n(\ttz)$,  
equals $\phi(n)$.

\begin{prpstn} \label{prof:une-mat-deux}
Let $i$  denote the imaginary unit and let $\zeta \defeq  \frac{1}{2} + i \frac{\sqrt{3}}{2}$:
\begin{itemize}
\item $\zeta$ and $\zeta^5 = \frac{1}{2} - i \frac{\sqrt{3}}{2}$ are the primitive sixth roots of unity,
  \item $i$ and $-i$ are the primitive fourth roots of unity, and 
\item $\zeta^2 = -\frac{1}{2} + i \frac{\sqrt{3}}{2}$ and $\zeta^4 = - \frac{1}{2} - i \frac{\sqrt{3}}{2}$ are the primitive cube roots of unity.  
\end{itemize}
For every  $M \in \carre{\Q}{2}$,
$M$ is torsion \textiff{} one of the following ten matrices is a Jordan normal form of $M$: 
\begin{center}
$\smalltwotwo
{0}{0}
{0}{0}$, 
$\smalltwotwo
{0}{1}
{0}{0}$,
$\smalltwotwo
{1}{0}
{0}{0}$,
$\smalltwotwo
{-1}{0}
{ 0}{0}$,
$\smalltwotwo
{1}{0}
{0}{1}$,
$\smalltwotwo
{-1}{ 0}
{ 0}{-1}$,
$\smalltwotwo
{1}{ 0}  
{0}{-1}$,
$\smalltwotwo
{\zeta^2}{ 0 }
{0}{\zeta^4}$, 
$\smalltwotwo
{i}{ 0}  
{0}{-i}$, 
or 
$\smalltwotwo
{\zeta}{ 0}  
{0}{\zeta^5}$.
\end{center}
\end{prpstn}

\begin{proof}
It is easy to check that the ten matrices listed above are torsion, 
so the ``if part''  holds true.
Let us now prove the ``only if part''.

Assume that $M$ is torsion.
Theorem~\ref{caract-torsion-C} implies that $M$ is nilpotent or diagonalizable (over $\C$).
If $M$ is nilpotent then either $M$ equals $\smalltwotwo{0}{0}{0}{0}$,
 or 
$\smalltwotwo{0}{1}{0}{0}$ is the Jordan normal form of  $M$.
Hence, we may assume that $M$ is diagonalizable for the rest of the proof.
Let $\chi(\ttz)$ denote the characteristic polynomial of $M$.

First, assume that $\chi(\ttz)$ is reducible over $\Q$.
Since $\chi(\ttz)$ is of degree $2$, the eigenvalues of $M$ are rational numbers.
Besides, Theorem~\ref{caract-torsion-C} implies that 
every non-zero eigenvalue of $M$ is a root of unity.
Since $-1$ and $+ 1$ are the only rational roots of unity, the eigenvalues of $M$ lie in the set  $\{ - 1, 0, + 1 \}$.
Hence, one of the following six matrices is a Jordan normal form of $M$:
$\smalltwotwo
{0}{0}
{0}{0}$, 
$\smalltwotwo
{1}{0}
{0}{0}$,
$\smalltwotwo
{-1}{0}
{ 0}{0}$,
$\smalltwotwo
{1}{0}
{0}{1}$,
$\smalltwotwo
{-1}{ 0}
{ 0}{-1}$,
or 
$\smalltwotwo
{1}{ 0}  
{0}{-1}$.

Second, assume that $\chi(\ttz)$ is irreducible over $\Q$.
Then, $\chi(\ttz)$ is a cyclotomic polynomial. 
Besides, it follows from Lemma~\ref{lem:Euler-naif} that the only cyclotomic polynomials of degree $2$ are: 
\begin{enumerate} 
\item $\Phi_3(\ttz) = \ttz^2 + \ttz + 1 = (\ttz - \zeta^2)(\ttz - \zeta^4)$, 
\item $\Phi_4(\ttz) = \ttz^2 + 1 = (\ttz - i)(\ttz + i)$, and 
\item $\Phi_6(\ttz) = \ttz^2 - \ttz + 1 = (\ttz - \zeta)(\ttz - \zeta^5)$. 
\end{enumerate}
Therefore, one of the following three matrices is a Jordan normal form of $M$:
$\smalltwotwo
{\zeta^2}{ 0 }
{0}{\zeta^4}$,
$\smalltwotwo
{i}{0}  
{0}{-i}$,
or
$\smalltwotwo
{\zeta}{0}  
{0}{\zeta^5}$.
\end{proof}

Note that 
$\smalltwotwo{\zeta^2}{0}{0}{\zeta^4}$,
$\smalltwotwo{i}{0}{0}{-i}$, and 
$\smalltwotwo{\zeta}{0}{0}{\zeta^5}$
are the Jordan normal forms of the integer matrices 
$\smalltwotwo{0}{-1}{1}{-1}$,
$\smalltwotwo{0}{-1}{1}{0}$, and
$\smalltwotwo{0}{-1}{1}{1}$,
respectively.

\subsection{The matrix torsion problem}
\label{sec:mat-tor-pb}

\begin{dfntn}
Define the \pbtorsion{} problem as: 
given an integer $d \ge 1$ and a matrix $M \in \carre{\Q}{d}$, 
decide whether $M$ is torsion. 
\end{dfntn}

For every integer $d \ge 1$, 
the complementary problem of \pbkfree{1}{\carre{\Q}{d}} is a restriction of \pbtorsion.
The main aim of this section is to prove: 

\begin{thrm} \label{th:torsion-polynomial}
The \pbtorsion{} problem is decidable in polynomial time.
\end{thrm}

The decidability of \pbtorsion{}  is an immediate corollary of the following useful result:

\begin{thrm}[Mandel and Simon  \cite{MandelS77}] \label{th:MandelSimon}
There exists a computable function $r\colon \N \setminus \{ 0 \} \to \N \setminus \{ 0 \}$ such that 
for every integer $d \ge 1$ and every matrix $M \in \carre{\Q}{d}$, 
$M^d = M^{d + r(d)}$ \textiff{} $M$ is torsion. 
\end{thrm}

\begin{proof}[Sketch of proof]
 Let $\phi$ denote Euler's totient function.
For each $d \in \N$, define $r(d)$ as the least common multiple of those $n \in \N$ for which $\phi(n) \le d$.
\end{proof}

In addition to the decidability of \pbtorsion{}, it follows from Theorem~\ref{th:MandelSimon} that,
for each integer $d \ge 1$,
\pbkfree{1}{\carre{\Q}{d}} is decidable in polynomial time.
However, 
$r$ is not polynomially bounded,
 so Theorem~\ref{th:torsion-polynomial} is not clear yet.

\begin{dfntn}[The matrix power problem]
For each $k \in \N$, 
define \pbkmatpow{k} as the following problem: 
given an integer $d \ge 1$ and two  matrices $A$, $B \in \carre{\Q}{d}$,
decide whether there exists $n \in \N$ such that $A = B^{n+k}$. 
\end{dfntn}

\begin{thrm}[Kannan and Lipton \cite{KannanL86}] \label{th:KannanLipton}
The \pbkmatpow{0} problem is decidable in polynomial time. 
\end{thrm}

\begin{crllr} \label{cor:MPk}
For every $k \in \N$,  \pbkmatpow{k} is decidable in polynomial time.  
\end{crllr}

\begin{proof}
By Theorem~\ref{th:KannanLipton}, 
it suffices to show that there exists a polynomial-time many-one reduction from \pbkmatpow{k}  to \pbkmatpow{0}.
Let $O_k$ denote the $k$-by-$k$ zero matrix.
Let $N_k$ denote the $k$-by-$k$ matrix defined by:
for all indices  $i$, $j \in \seg{1}{k}$, 
the ${(i, j)}^\text{th}$ entry of $N_k$ equals one if $j - i =  1$,
and zero otherwise.
For instance, in the case where $k = 4$, we have 
\begin{align*}
N_4 & =
\begin{bmatrix}
 0 & 1 & 0 & 0 \\
 0 & 0 & 1 & 0  \\
0 & 0 & 0 & 1  \\
0 & 0 & 0 & 0  \\
 \end{bmatrix} \, , 
&   
N^2_4 & =
\begin{bmatrix}
0 & 0 & 1 & 0 \\
0 & 0 & 0 & 1  \\
0 & 0 & 0 & 0  \\
0 & 0 & 0 & 0  \\
 \end{bmatrix}  \, ,
& &  
\text{and} 
&  
N_4^3  & =
\begin{bmatrix}
0 & 0 & 0 & 1 \\
0 & 0 & 0 & 0  \\
0 & 0 & 0 & 0  \\
0 & 0 & 0 & 0  \\
 \end{bmatrix} \, .
\end{align*}
It is easy to see that, for every $n \in \N$, 
$N_k^n = O_k$ is equivalent to $n \ge k$ 
($N_k^{k - 1}$ has a one in its right-upper corner).

Let $(d, A, B)$ be an instance  of \pbkmatpow{k}.
Define two matrices $C$, $D \in \carre{\Q}{(k + d)}$ by:
\begin{align*}
C & \defeq \twotwo{A}{O}{O}{O_k}
& & \text{and} &
D & \defeq \twotwo{B}{O}{O}{N_k} \, .
\end{align*}
Clearly, 
$(k + d, C, D)$ is an instance of  \pbkmatpow{0}
and 
$(k + d, C, D)$  is computable from $(d, A, B)$ in polynomial time. 
For every $n \in \N$,
$C = D^n$ is equivalent to the conjunction of $A = B^n$ and $n \ge k$.
Therefore, $(d, A, B)$ is a yes-instance of \pbkmatpow{k} \textiff{} $(k + d, C, D)$ is a yes-instance of \pbkmatpow{0}. 
\end{proof}

\begin{rmrk} \label{rem:semmem1}  
For every integer $d \ge 1$, \pbksemmem{1}{\carre{\Q}{d}} can be seen as a restriction of \pbkmatpow{1}, 
so it follows from Corollary~\ref{cor:MPk} that \pbksemmem{1}{\carre{\Q}{d}} is decidable in polynomial time. 
\end{rmrk}

\begin{proof}[Proof of Theorem~\ref{th:torsion-polynomial}]
Let $(d, M)$ be an instance of \pbtorsion.
It follows from Theorem~\ref{caract-torsion-C} that
$(d, M)$ is a yes-instance of \pbtorsion{} 
\textiff{} 
$(d, M^d, M^d)$ is a yes-instance \pbkmatpow{2}.
Hence, there exists a polynomial-time many-one reduction from \pbtorsion{} to \pbkmatpow{2}, 
and thus \pbtorsion{} is decidable in polynomial time by Corollary~\ref{cor:MPk}.
\end{proof}

At this point, it is interesting to briefly  discuss about power boundedness.

\begin{dfntn}
A complex square matrix $M$ is called \emph{power bounded} if the semigroup $\left\{ M, M^2, M^3, M^4, \dotsc \right\}$ is bounded. 
Define the \pbmpb{} problem as: 
given an integer $d \ge 1$ and a matrix $M \in \carre{\Q}{d}$, decide whether $M$ is power bounded. 
\end{dfntn}

For every integer $d \ge 1$,  \pbkbound{1}{\carre{\Q}{d}} is a restriction of \pbmpb.

\begin{prpstn} \label{prop:power-bound}
The \pbmpb{} problem is decidable. 
\end{prpstn}

\begin{proof}
Let $M$ be a complex square matrix.
Let $\mu(\ttz)$ denote the minimal polynomial of $M$:
the  roots of $\mu(\ttz)$ are the eigenvalues of $M$.
Put $\nu(\ttz) \defeq \gcd(\mu(\ttz), \mu'(\ttz))$: the roots of $\nu(\ttz)$ are the multiple roots of $\mu(\ttz)$.
It is easy to see that $M$ is power bounded \textiff{} the following two conditions are met:
\begin{enumerate}[$(i)$.]
 \item every root of $\mu(\ttz)$ has modulus at most $1$ and
 \item every root of $\nu(\ttz)$ has modulus less than $1$.
\end{enumerate}

Now, consider the case where every entry of $M$ is in $\Q$.
Then, $\mu(\ttz)$  and $\nu(\ttz)$ are computable from $M$ in polynomial time \cite{KannanL86}.
Therefore, deciding whether $M$ is power bounded reduces to checking conditions~$(i)$ and $(ii)$.
This can be achieved using Tarski's decision procedure \cite{Tarski51}.
\end{proof}

We conjecture that  \pbmpb{} is decidable in polynomial time.

%
%
%

\subsection{The morphism torsion problem} \label{sec:mortor}

\begin{dfntn} \label{def:hom-Sigma}
For any alphabet $\Sigma$,
let $\hom(\Sigma^\star)$ denote the set of all morphisms from $\Sigma^\star$ to itself.
Define the \pbmortor{} problem as: 
given a finite alphabet $\Sigma$ and a morphism $\sigma \in \hom(\Sigma^\star)$, decide whether $\sigma$ is torsion (under function composition).
\end{dfntn} 

The size of an instance $(\Sigma, \sigma)$ of \pbmortor{} equals  $\sum_{a \in \Sigma} \left(1 + \lgr{\sigma(a)} \right)$.

\begin{dfntn}[Incidence matrix] \label{def:inc-mat}
Let $\Sigma$ be a finite alphabet, 
let $d$ denote the cardinality of $\Sigma$, and 
let $a_1$, $a_2$, \dots, $a_d$ be such that $\Sigma = \{ a_1, a_2, \dotsc, a_d \}$:
$a_1a_2 \dotsb a_d$ is a permutation of $\Sigma$.
The \emph{incidence matrix} of $\sigma$ relative to $a_1a_2 \dotsm a_d$ is defined as
$$
\begin{bmatrix}
 \nocc{a_1}{\sigma(a_1)} &  \nocc{a_1}{\sigma(a_2)} & \cdots & \nocc{a_1}{\sigma(a_d)} \\
 \nocc{a_2}{\sigma(a_1)} &  \nocc{a_2}{\sigma(a_2)} & \cdots & \nocc{a_2}{\sigma(a_d)} \\
\vdots & \vdots & \ddots & \vdots \\
\nocc{a_d}{\sigma(a_1)} &  \nocc{a_d}{\sigma(a_2)} & \cdots & \nocc{a_d}{\sigma(a_d)} \\
\end{bmatrix} \, .
$$
\end{dfntn}
The incidence matrix of $\sigma$ relative to $a_1a_2 \dotsm a_d$ belongs to $\carre{\N}{d}$;
for all $i$, $j \in \seg{1}{d}$, its $(i, j)^\text{th}$ entry equals the number of occurrences of $a_i$ in $\sigma(a_j)$.

\begin{clm} \label{claim:mat-morph}
Let $\Sigma$ be a finite alphabet.
For each $\sigma \in \hom(\Sigma^\star)$, let $P_\sigma$ denote the incidence matrix of $\sigma$ relative to some fixed permutation of $\Sigma$.
\begin{enumerate}[$(i)$.]
\item 
Equality $P_{\sigma} P_{\tau} = P_{\sigma  \tau}$ holds for all $\sigma$, $\tau \in \hom(\Sigma^\star)$.
\item 
For each $P \in \carre{\N}{d}$, there exist at most finitely many $\tau \in \hom(\Sigma^\star)$ such that $P_\tau = P$.
\end{enumerate}
\end{clm}

\begin{thrm} \label{th:mortor}
The \pbmortor{} problem is decidable in polynomial time.
\end{thrm}

\begin{proof}
By Theorem~\ref{th:torsion-polynomial}, 
it suffices to show that there exists a polynomial-time many-one reduction from  \pbmortor{} to \pbtorsion.
The idea is to prove that a morphism is torsion \textiff{} its incidence matrix is torsion.

Let $(\Sigma, \sigma)$ be an instance of \pbmortor.
Let $d$ denote the cardinality of $\Sigma$.
For each $\tau \in \hom(\Sigma^\star)$, 
let $P_\tau$ denote the incidence matrix of $\tau$ relative to some fixed permutation of $\Sigma$.
Clearly, 
$(d, P_\sigma)$ is an instance of \pbtorsion{} 
and 
$(d, P_\sigma)$ is computable from $(\Sigma, \sigma)$ in polynomial time.

Let us check that $(\Sigma, \sigma)$ is a yes-instance of \pbmortor{} \textiff{} $(d, P_\sigma)$ is a yes-instance of \pbtorsion.
It follows from Claim~\ref{claim:mat-morph}.$(i)$ that $P^n_\sigma = P_{\sigma^n}$ for every $n \in \N$.   
Therefore, if $\sigma$ is torsion then $P_\sigma$ is torsion.
Conversely, 
assume that $P_\sigma$ is torsion.
Then, the set of matrices 
$\calp \defeq \{ P_\sigma, P_\sigma^2, P_\sigma^3, P_\sigma^4, \dotsc \}$ is finite,
so by Claim~\ref{claim:mat-morph}.$(ii)$, there exist at most finitely many $\tau \in \hom(\Sigma^\star)$  such that $P_\tau \in \calp$. 
Since  $P_{\sigma^n} \in \calp$ for every integer $n \ge 1$, 
the set $\left\{ \sigma, \sigma^2, \sigma^3, \sigma^4, \dotsc \right\}$ is finite, and thus $\sigma$ is torsion.
\end{proof}

\begin{crllr}
For any finite  alphabet $\Sigma$, 
\pbkfree{1}{\hom(\Sigma^\star)} is decidable in polynomial time.
\end{crllr}

\begin{open}[Richomme \cite{RichommePriv07}]  
For any finite alphabet $\Sigma$ with cardinality greater than $1$ and any integer $k > 1$, 
the decidability of \pbkfree{k}{\hom(\Sigma^\star)} is open.
\end{open}

The decidability of \pbkfree{2}{\hom(\FreeM)} is tackled in Section~\ref{sec:two-two-morph}.

\section{Balanced equations} \label{sec:balanced-eq}

This section centers on the consequences of the following lemma:

\begin{lmm} \label{lem:eq:homo}
Let $S$ be a semigroup and let $X$ be a subset of $S$ with cardinality greater than $1$.
The set $X$ is not a code 
\textiff{} 
there exist $x$, $x' \in X$ and $z$, $z' \in X^+$ such that $x \ne x'$ and $z x z x' z' = z x' z' x z$.
\end{lmm}

\begin{proof}
The ``if part'' is clear. Let us prove the ``only if part''.

Let $\Sigma$ be an alphabet and let $\sigma\colon \Sigma^+ \to S$ be a morphism such that $\sigma$ induces a bijection from $\Sigma$ onto $X$.
Assume that $X$ is not a code.
By  Claim~\ref{clm:pre-image-code}, $\sigma$ is non-injective.
Therefore, 
there exist $w$, $w' \in \Sigma^+$ such that $w \ne w'$ and $\sigma(w) = \sigma(w')$.

 First, assume that $w$ is not a prefix of $w'$ and that $w'$ is not a prefix of $w$.
Then, there exist $a$, $a' \in \Sigma$ and $u$, $v$, $v' \in \Sigma^\star$ such that 
$a \ne a'$,
$w = u a v$, and 
$w' = u a' v'$: 
$u$ is the longest common prefix of $w$ and $w'$.
Note that  $u$, $v$, or $v'$ may be the empty word.
It is easy to see that 
$\sigma(a)$, $\sigma(a')$, $\sigma(v a u)$, and $\sigma(v' a u)$
are suitable choices for
$x$, $x'$, $z$, and $z'$, respectively. 

Second, assume that $w$ is a proper prefix of $w'$. 
Then, there exists $a \in \Sigma$ such that $wa$ is a prefix of $w'$.
Since $\Sigma$ and $X$ are equinumerous,
the cardinality of $\Sigma$ is greater than $1$,
and thus there exists $b \in \Sigma$ such that $a \ne b$.
Clearly, we have 
$\sigma(w b) = \sigma(w'b)$,
$w b$ is not a prefix of $w' b$, and 
$w' b$ is not a prefix of $w b$.
Therefore, the second case reduces to the first case.

Third, assume that $w'$ is a proper prefix of $w$.
Since $w$ and $w'$ play symmetric roles, the third case reduces to the second case.
\end{proof}

For each  $y \in X^+$, 
the factorizations of $y$ over $X$ are in one-to-one correspondence with the preimages of $y$ under $\sigma$.
Let $\bar x$, $\bar x' \in \Sigma$ and $\bar z$, $\bar z' \in \Sigma^+$ be such that 
$x = \sigma(\bar x)$,
$x' = \sigma(\bar x')$,
$z = \sigma(\bar z)$, and 
$z' = \sigma(\bar z')$.
Equation $ z x z x' z' = z x' z' x z$ is ``balanced'' in the sense that 
the word $\bar z \bar x \bar z \bar  x'  \bar z'$, 
which corresponds to a factorization of the left-hand side,
is a permutation of the word $\bar z \bar x' \bar z'\bar x \bar z$,
which corresponds to a factorization of the right-hand side.

\subsection{Cancellation}

\begin{dfntn}[Cancellation]
Let $S$ be a semigroup and let $s \in S$.
We say that $s$ is \emph{left-cancellative} in $S$ if for all $u$, $v \in S$, $s u = s v$ implies $u = v$.
In the same way, we say that
$s$ is \emph{right-cancellative} in $S$ if for all $u$, $v \in S$, $u s = v s$ implies $u = v$.  
We say that $s$ is \emph{cancellative} in $S$ if $s$ is both left-cancellative and right-cancellative in $S$.
\end{dfntn}

\begin{xmpl} \label{ex:simplifiable}
Let $X$ be a (finite or  infinite) set and let $S$ denote the set of all functions from $X$ to itself.
Clearly, $S$ is a semigroup under function composition. 
The left-cancellative elements of $S$ are the injections,
the right-cancellative elements of $S$ are the surjections, 
and 
the cancellative elements of $S$ are the bijections.
\end{xmpl}

The first useful corollary of Lemma~\ref{lem:eq:homo} is:

\begin{lmm} \label{lem:eq-group}
Let $S$ be a semigroup and let $X$ be a subset of $S$ such that
the cardinality of $X$ is greater than $1$ and 
every element of $X$ is left-cancellative in $S$.
The set  $X$ is not a code \textiff{}
there exist $x$, $x' \in X$ and $z$, $z' \in X^+$ 
such that $x \ne x'$ and $x z = x'z'$.
\end{lmm}

\begin{proof}
 The ``if part'' is clear. It remains to prove the ``only if part''.

Assume that $X$ is not a code.
By Lemma~\ref{lem:eq:homo}, 
there exist $x$, $x' \in X$ and $t$, $t' \in X^+$ such that 
 $t x t x' t' = t x' t' x t$.
Since $t$ is left-cancellative, 
equality $x z = x'z'$ holds with $z \defeq  t x' t'$ and $z' \defeq  t' x t$.
\end{proof}

Lemma~\ref{lem:eq-group} is extensively used throughout the paper.
The following example shows that Lemma~\ref{lem:eq-group} does not hold without any cancellation property:

\begin{xmpl}
Let 
\begin{align*}
X & \defeq \twotwo{4}{2}{2}{1} & &  \text{and}  &  
X' & \defeq \twotwo{1}{2}{2}{4} \,.
\end{align*} 
The set $\{ X, X' \}$ is not a code under matrix multiplication because $XXX'X = XX'XX$.
Besides, the row matrix $L \defeq \onetwo{-1}{2}$ satisfies $L X = \onetwo{0}{0}$ and $L X' = \onetwo{3}{6}$.
For all $Z$, $Z' \in {\{ X, X' \}}^+$,
we thus have $L X Z = \onetwo{0}{0}$ while 
the entries of $L X' Z'$ are positive.
Therefore, $X Z$ and $X' Z'$ are distinct for all $Z$, $Z' \in {\{ X, X' \}}^+$.
\end{xmpl}

\subsection{Direct products of semigroups}

Given two semigroups $S$ and $T$ such that $T$ is commutative, 
let us  characterize those subsets of $S \times T$ that are codes.

\begin{lmm} \label{lem:code-produit}
Let $S$ and $T$ be two semigroups and let $Z$ be a subset of $S \times T$  such that 
$T$ is commutative 
and 
the cardinality of $Z$ is greater than $1$.
Let $\alpha\colon S \times T \to S$ be defined by:
$\alpha(s, t) \defeq s$ for every $(s, t) \in S \times T$.
The set $Z$ is a code \textiff{} the following two conditions are met:
$\alpha$ is injective on $Z$ and
$\alpha(Z)$ is a code.
 \end{lmm}

\begin{proof}
The ``if part'' follows from Claim~\ref{clm:pre-image-code}.
It remains to prove the ``only if part''.

First,
assume that $\alpha$ is non-injective on $Z$.
Then, there  exist $x \in S$ and $y$, $y' \in T$ such that 
$y \ne y'$,
$(x, y) \in Z$, and
$(x, y') \in Z$.  
Since $(x, y)$ and $(x, y')$ commute, $Z$ is not a code.

Second, assume that $\alpha(Z)$ is not a code.
By Lemma~\ref{lem:eq:homo}, 
there exist 
$(x, y)$, $(x', y') \in Z$ 
and 
$(u, v)$, $(u', v') \in Z^+$ such that 
$x \ne x'$ and $u x u x' u' = u x' u' x u$.
Now, remark that $v y v y' v' = v y' v' y v$ because $T$ is commutative.
Hence, we have
$$
(u, v) (x, y) (u, v) (x', y') (u', v') 
 = (u, v) (x', y') (u', v') (x, y) (u, v)  \, , 
$$
and thus $Z$ is not a code.
\end{proof}

To complete Lemma~\ref{lem:code-produit}, let us characterize those elements of $S \times T$ that are torsion: 
for every $(s, t) \in S \times T$, $(s, t)$ is torsion \textiff{} both $s$ and $t$ are torsion.

\begin{lmm} \label{lem:Xy-X}
Let $S$ and $T$ be two semigroups and let $y \in T$. 
For every subset $X \subseteq S$ such that the cardinality of $X$ is greater than $1$,
$X \times \{ y \}$ is a code \textiff{} $X$ is a code.
\end{lmm}

\begin{proof}
Although $T$ is not necessarily commutative, 
$T' \defeq \left\{ y, y^2, y^3, y^4, \dotsc\right\}$ is a commutative subsemigroup of $T$ such that $X \times \{ y \} \subseteq S \times T'$.
The desired result can thus be deduced from Lemma~\ref{lem:code-produit}.
\end{proof}

\begin{thrm} \label{th:sg-et-prod}
Let $S$ and $T$ be two non-empty semigroups with \ruset{}s and let $k$ be an integer greater than $1$.
If \pbkfree{k}{S \times T} is decidable
then 
both  \pbkfree{k}{S} and  \pbkfree{k}{T} are decidable.
\end{thrm} 

\begin{proof}
Let $y$ be a fixed element of $T$.
For each $k$-element subset $X \subseteq S$,  
$X \times \{ y \}$ is a $k$-element subset of $S \times T$, 
and according to Lemma~\ref{lem:Xy-X}, 
$X$ is a code \textiff{} $X \times \{ y \}$ is a code.
Hence, 
there exists a many-one reduction from \pbkfree{k}{S} to \pbkfree{k}{S \times T}.
In the same way,
\pbkfree{k}{T} reduces to \pbkfree{k}{S \times T}.
\end{proof}

The converse of Theorem~\ref{th:sg-et-prod} is false in general: 
for instance, 
\pbfree{\FreeM} is decidable (see Example~\ref{ex:sardine})
while 
\pbkfree{k}{\zeonzeon} is undecidable for every integer $k \ge 13$ (see Section~\ref{sec:three-three}).  
An interesting partial converse is:

\begin{lmm} \label{lem:sg-et-comm}
Let $S$ and $T$ be two semigroups with \ruset{}s and let $k$ be an integer greater than $1$.
If  \pbkfree{k}{S} is decidable and if $T$ is commutative then \pbkfree{k}{S \times T} is decidable.
\end{lmm} 

\begin{proof}
It follows from Lemma~\ref{lem:code-produit} that \pbkfree{k}{S \times T} reduces to \pbkfree{k}{S}.
\end{proof}

 Theorem~\ref{th:sg-et-prod} and Lemma~\ref{lem:sg-et-comm} deserve further comments.
By Theorem~\ref{th:pathos} below, 
there exists a commutative (semi)group $T$  with a \ruset{} such that \pbkfree{1}{T} is undecidable.
However,  \pbkfree{1}{{{\{ \on \}}^+ \times T}} is decidable because no element of ${\{ \on \}}^+ \times T$ is torsion.
Hence, it is essential to assume  $k > 1$ in Theorem~\ref{th:sg-et-prod}. 
Moreover,  \pbkfree{1}{{\{ \mv \} \times T}} is undecidable while \pbkfree{1}{\{ \mv \}} is decidable, 
so it is also essential to assume $k > 1$ in Lemma~\ref{lem:sg-et-comm}.

\begin{dfntn}
For each $d \in \N$,
$\FreeM^{\times d}$ denotes the semigroup obtained as the direct product of $d$ copies of $\FreeM$: 
$\FreeM^{\times 0} = \{ \mv \}$, 
$\FreeM^{\times 1} = \FreeM$, 
$\FreeM^{\times 2} = \FreeM \times \FreeM$, 
$\FreeM^{\times 3} = \FreeM \times \FreeM \times \FreeM$, 
\emph{etc}. 
\end{dfntn}

\begin{thrm} 
Let $n$ be a positive integer and let $\Sigma_1$, $\Sigma_2$, \ldots, $\Sigma_n$ be $n$ finite alphabets.
Let $d$ denote the number of $i \in \seg{1}{n}$ such that the cardinality of $\Sigma_i$ is greater than $1$.
For every integer $k \ge 1$, 
\pbkfree{k}{\Sigma_1^\star \times \Sigma_2^\star \times \dotsb \times  \Sigma_n^\star} is decidable 
\textiff{}  
\pbkfree{k}{\FreeM^{\times d}} is decidable.
\end{thrm} 

\begin{proof}
We only need to consider the case where $k > 1$ because both problems are trivially decidable in the case where $k = 1$.
Moreover, for each permutation $(i_1, i_2, \dotsc, i_n)$ of $\seg{1}{n}$,
$\Sigma_1^\star \times \Sigma_2^\star \times \dotsb \times  \Sigma_n^\star$  
and 
$\Sigma_{i_1}^\star \times \Sigma_{i_2}^\star \times \dotsb \times  \Sigma_{i_n}^\star$
are isomorphic:
the function mapping each 
$(w_1, w_2, \dotsc, w_n) \in \Sigma_1^\star \times \Sigma_2^\star \times \dotsb \times  \Sigma_n^\star$ 
to 
$\left(w_{i_1}, w_{i_2}, \dotsc, w_{i_n} \right)$ is an isomorphism.
Therefore, 
we may assume without loss of generality that the cardinality of $\Sigma_i$ is greater than $1$ for every $i \in \seg{1}{d}$, 
or equivalently, 
that $\Sigma_i^\star$ is commutative for every $i \in \seg{d + 1}{n}$.
Hence, 
it follows from Lemma~\ref{lem:sg-et-comm} that
\pbkfree{k}{\Sigma_1^\star \times \Sigma_2^\star \times \dotsb \times  \Sigma_n^\star} is decidable \textiff{} 
\pbkfree{k}{S} is decidable, where $S \defeq \Sigma_1^\star \times \Sigma_2^\star \times \dotsb \times  \Sigma_d^\star$.

For each $i \in \seg{1}{d}$, 
let $\phi_i\colon \FreeM \to \Sigma_i^\star$ be an injective morphism, 
\emph{e.g.},
 $\phi_i$ can be any  morphism extending an injection from $\zeon$ to $\Sigma_i$.
The function mapping 
each
$(u_1, u_2, \dotsc, u_d) \in  \FreeM^{\times d}$ 
to 
$\left(\phi_1(u_1), \phi_2(u_2), \dotsc, \phi_d(u_d) \right)$
is an injective morphism from $\FreeM^{\times d}$ to 
$S$;
it induces a one-one reduction from 
\pbkfree{k}{\FreeM^{\times d}}
to 
\pbkfree{k}{S}.
Hence, the ``only if part'' of the theorem holds.

For each $i \in \seg{1}{d}$, 
let $\psi_i\colon \Sigma_i^\star \to \FreeM$ be an injective morphism (see Claim~\ref{claim:inject-zeon}).
The function mapping 
each
$(v_1, v_2, \dotsc, v_d) \in S$ 
to 
$(\psi_1(v_1), \psi_2(v_2), \dotsc, \psi_d(v_d))$
is an injective morphism from $S$ to $\FreeM^{\times d}$;
it induces a one-one reduction from 
\pbkfree{k}{S}
to
\pbkfree{k}{ \FreeM^{\times d}}.
Hence, the ``if part'' of the theorem holds.
\end{proof}

\subsection{Rational matrices versus integer matrices}
\label{sec:rat-int}

The following lemma generalizes Lemma~3 in \cite{CassaigneHK99}:

\begin{lmm} \label{lem:mult-lambda}
Let $d$ be a positive integer,
let $\calx$ be a subset of $\carre{\C}{d}$ with cardinality greater than $1$, and 
let $\lambda\colon \calx \to \C \setminus \{ 0 \}$.
The set $\calx$ is a code under matrix multiplication \textiff{} the following two conditions are met:
\begin{enumerate}[$(i)$.]
 \item $\left\{ \lambda(X) X : X \in \calx \right\}$ is a code under matrix multiplication  and 
\item for all $X$, $Y \in \calx$, $X \ne Y$ implies $\lambda(X) X \ne \lambda(Y) Y$.
\end{enumerate}
 \end{lmm}

\begin{proof}
Let $\calz \defeq \left\{ \left(X, \lambda(X) \right) : X \in \calx \right\}$.
By Lemma~\ref{lem:code-produit}, 
$\calz$ is a code \textiff{} $\calx$ is a code.
Let 
$\calz' \defeq \left\{ \left(\lambda(X) X, \lambda(X) \right) : X \in \calx \right\}$.
By Lemma~\ref{lem:code-produit}, 
$\calz'$ is a code \textiff{} conditions~$(i)$ and~$(ii)$ are met.
Let $\check \C \defeq  \C \setminus \{ 0 \}$.
Let $\sigma\colon \carre{\C}{d}  \times \check \C  \to \carre{\C}{d} \times \check \C$ be defined by:
$\sigma(X, a) \defeq (aX, a)$ for every $(X, a) \in \carre{\C}{d} \times \check \C$.
Remark that
$\sigma(\calz) = \calz'$,
$\sigma$ is injective, and 
$\sigma$ is a morphism:
$\sigma(XY, ab) = ((aX )(bY), ab) = \sigma(X, a)\sigma(Y, b)$
for all $X$, $Y \in \carre{\C}{d}$ and all $a$, $b \in \check \C$.
Therefore, 
$\calz$ is code \textiff{} $\calz'$ is a code. 
We have thus proven that the following four assertions are equivalent: 
$\calx$ is a code,
$\calz$ is a code,
$\calz'$ is a code, and
conditions~$(i)$ and~$(ii)$ are met.
\end{proof}

Let 
$X \defeq \smalltwotwo{2}{0}{0}{2}$, 
$X' \defeq \smalltwotwo{1}{0}{0}{1}$, 
$\calx \defeq \left\{ X, X' \right\}$,
$\lambda(X) \defeq 1$, and
$\lambda(X') \defeq 2$.
Note that $\lambda(X) X = \lambda(X')X'$.
Clearly,
$\left\{ \lambda(X) X, \lambda(X') X'  \right\} = \{ X \}$ is a code under matrix multiplication 
but $\calx$ is not.
Hence, condition~$(ii)$ is crucial in  Lemma~\ref{lem:mult-lambda}.

The main result of the section is now easy to prove: 

\begin{thrm} \label{th:Z-Q}
For all integers $k$, $d \ge 1$,
\pbkfree{k}{\carre{\Q}{d}} is decidable 
\textiff{} 
\pbkfree{k}{\carre{\Z}{d}} is decidable.
\end{thrm}

\begin{proof}
The ``only if part'' holds because \pbkfree{k}{\carre{\Z}{d}} is a restriction of \pbkfree{k}{\carre{\Q}{d}}.
Moreover, 
\pbkfree{1}{\carre{\Q}{d}} is decidable by Theorems~\ref{th:MandelSimon} or  \ref{th:torsion-polynomial}.
Hence, to conclude the proof of the theorem, 
it suffices to show that there exists a many-one reduction from  \pbkfree{k}{\carre{\Q}{d}} to  \pbkfree{k}{\carre{\Z}{d}} in the case where $k > 1$.

For each finite subset $\calx \subseteq \carre{\Q}{d}$,
let $t(\calx)$ denote the smallest  integer $n \ge 1$ such that $n X \in \carre{\Z}{d} $ for every $X \in \calx$.
For each instance $\calx$ of \pbkfree{k}{\carre{\Q}{d}},
$\calx' \defeq \left\{ t(\calx) X : X \in \calx \right\}$ is an instance of \pbkfree{k}{\carre{\Z}{d}}, 
$ \calx'$ is computable from $\calx$, 
and by Lemma~\ref{lem:mult-lambda},
$\calx$ is a yes-instance of \pbkfree{k}{\carre{\Q}{d}} 
\textiff{}  
$\calx'$ is a yes-instance of \pbkfree{k}{\carre{\Z}{d}}. 
\end{proof}

To conclude the section, 
let us discuss whether analogues of Theorem~\ref{th:Z-Q} hold for mortality, boundedness, and semigroup membership.
First, for all integers $k$, $d \ge 1$,
\pbkmort{k}{\carre{\Q}{d}} is decidable 
\textiff{} 
\pbkmort{k}{\carre{\Z}{d}} is decidable:
for every $k$-element subset $\calx \subseteq \carre{\Q}{d}$, 
$\calx$ is a yes-instance of \pbkmort{k}{\carre{\Q}{d}} 
\textiff{}
$\calx'$ is yes-instance \pbkmort{k}{\carre{\Z}{d}}, 
where $\calx'$ is as in the proof of Theorem~\ref{th:Z-Q}.
Note in passing that the decidability of \pbkmort{k}{\carre{\Q}{d}} is still open for several pairs $(d, k)$ of positive integers \cite{Claus07,BB02}. 
Second, \pbkbound{2}{\carre{\Q}{47}} is undecidable \cite{BlondelC03},
but for every integer $d \ge 1$,  \pbbound{\carre{\Z}{d}}  is decidable: 
\pbbound{\carre{\Z}{d}} is in fact the same problem as \pbsemfin{\carre{\Z}{d}} and 
the latter is decidable \cite{MandelS77,Jacob77}.
Third, it is still unknown whether there exist positive integers $k_0$ and $d_0$ satisfying the following two properties:
\pbksemmem{k_0}{\carre{\Q}{d_0}} is undecidable and \pbksemmem{k_0}{\carre{\Z}{d_0}} is decidable.

\section{Subsemigroups of groups} \label{sec:semigroup-group}

Automata over monoids, and in particular automata over the free group, have been widely studied \cite{Saka03,Grunschlag99}.
In this section, we first prove that for any group $G$ with a \ruset, 
\pbfree{G} reduces to an automata theory problem (Theorem~\ref{th:ramemb-freeness-gen}).
We then use this reduction to show that both $\GL{2}{\Z}$ and the free groups have decidable freeness problems
(Corollaries~\ref{coro:GL2} and~\ref{coro:free-group}).

\begin{rmrk}
Let $G$ be a group with a \ruset.
\pbfree{G} should not be confused with the following problem,
which is not the concern of the paper:
given a finite subset  $X \subseteq G$, decide whether the subgroup of $G$ generated by $X$ is a free group with basis $X$.
\end{rmrk} 

\begin{dfntn}[Automaton] \label{def:automate}
Let $X$ be a set.
An  \emph{automaton} over $X$ is a quadruple $A = (Q, E, I, T)$, where 
$Q$ is a set, 
$I$ and $T$ are subsets of $Q$,
and $E$ is a subset of $Q \times X \times Q$.
The elements of $Q$ are the \emph{states} of $A$,
the elements of $E$ are the \emph{transitions} of $A$,
the elements of $I$ are the \emph{initial} states of $A$, and
the elements of $T$ are the \emph{terminal} states of $A$.
We say that $A$ is \emph{finite} if $Q$ and $E$ are finite.
A transition $(p, s, q) \in E$ is usually denoted $\transit{p}{s}{q}$.
\end{dfntn}

Finite automata over finite alphabets play a central role in theoretical computer science;
they are termed ``nondeterministic automata'' or simply ``automata'' in most of the literature.
According to our definition,
an automaton over $X$ is also an automaton over any superset of $X$.
In particular,
for any alphabet $\Sigma$,
an automaton over $\Sigma$ is also an automaton over the free monoid $\Sigma^\star$.

\begin{dfntn}[Acceptance] \label{def:accept}
Let $M$ be a monoid, 
let $A$ be an automaton over $M$,
and 
let $s$ be an element  of $M$.
We say that $A$ \emph{accepts} $s$ if for some integer $n \in \N$, 
there exist $n + 1$ states $q_0$, $q_1$, \ldots, $q_n$ and $n$ elements $s_1$, $s_2$, \ldots, $s_n \in M$ meeting the following requirements:
$s = s_1 s_2 \dotsm s_n$, 
$q_0$ is an initial state of $A$,
$q_n$ is a terminal state of $A$, 
and  
$\transit{q_{i - 1}}{s_i}{q_i}$ is a transition of $A$ for every  $i \in \seg{1}{n}$.
Define the \emph{behavior} of $A$ as the set of those elements of $M$ that are accepted by $A$.
\end{dfntn} 

For every automaton $A = (Q, E, I, T)$  over $M$ such that  $I \cap T \ne \emptyset$, 
it follows from Definition~\ref{def:accept} that $A$ accepts the identity element of $M$.

Let $\Sigma$ be a finite alphabet.
Every  finite automaton over $\Sigma^\star$ can be transformed in polynomial time into a finite  automaton over $\Sigma \cup \{ \mv \}$ with the same behavior: 
simply split each transition labeled with a word of length greater than $1$.
Moreover, 
every finite automaton over $\Sigma \cup \{ \mv \}$ can be transformed in polynomial time into a finite  automaton over $\Sigma$ with the same behavior (Section~2.5 in \cite{HopcroftMU01}).

\begin{rmrk}[Kleene's theorem]
Let $M$ be a monoid.
 A subset of $M$ is called \emph{rational}  if it equals the behavior of some finite automaton over $M$.
We claim that the set of all rational subsets of $M$ equals the closure of the set of all finite subsets of $M$ under set union, 
set product, 
and star.
If $M = \Sigma^\star$ for some finite alphabet $\Sigma$ then our claim is simply Kleene's theorem (Section 3.2 in~\cite{HopcroftMU01}).  
The generalization to an arbitrary monoid is straightforward \cite{Saka03}.
\end{rmrk}

\begin{dfntn} 
For every monoid $M$ with a \ruset, 
define \pbratmem{M} as the following problem: 
given a finite automaton $A$ over $M$ and an element $s \in M$,
decide whether $A$ accepts $s$.
\end{dfntn}

Note that \pbratmem{M} is also known as the 
rational subset problem for $M$  \cite{KambitesSS07}
and 
as the rational membership problem over $M$ \cite{Grunschlag99}.

\begin{xmpl}[Section~4.3.3 in \cite{HopcroftMU01}]  \label{ex:sigmastar}
For any finite alphabet $\Sigma$,  
\pbratmem{\Sigma^\star}  is decidable in polynomial time.
\end{xmpl}

A \emph{group} is a monoid $G$ in which every element is invertible.  
The identity element of $G$ is denoted $\mvg$.
The \emph{inversion} in $G$ is the bijection function from $G$ onto itself that maps each element $g \in G$ to its inverse $g^{-1}$.
For every $x \in G$, $x$ is torsion \textiff{} there exists an integer $n \ge 1$ such that $x^n = \mvg$.

\begin{lmm} \label{lem:op-comput} \leavevmode
\begin{enumerate}[$(i)$.]
 \item Let $M$ be a monoid with a \ruset.
If \pbratmem{M} is decidable then the operation of $M$ is computable.
 \item Let $G$ be a group with a \ruset{}.
If the operation of $G$ is computable 
then the inversion in $G$ is computable.
\end{enumerate}
\end{lmm}

\begin{proof}
Let $x$, $y \in M$.
Let $A_{x, y}$ be the automaton over $\{ x,  y \}$ defined by:
\begin{itemize}
\item  $\ttI$, $\ttQ$, and $\ttT$ are the  states of $A_{x, y}$,
\item  $\transit{\ttI}{x}{\ttQ}$ and $\transit{\ttQ}{y}{\ttT}$ are the transitions of $A_{x, y}$,
\item $\ttI$ is the unique initial state of $A_{x, y}$, and
\item  $\ttT$ is the unique terminal state of $A_{x, y}$.
\end{itemize}  
Clearly, the behavior of $A_{x, y}$ equals $\{ xy \}$.
Now, assume that \pbratmem{M} is decidable.
To compute $xy$ from $x$ and $y$,
first compute $A_{x, y}$, 
and then examine the elements of $M$ one after another until finding the one that is accepted by $A_{x, y}$. 
We have thus proven part~$(i)$.

Let us now turn to part~$(ii)$.
Let $g$, $h \in G$.
To decide whether $h$ is the inverse of $g$, 
it suffice to compute $g h$ and then to test whether the result equals $\mvg$.
Hence, the inverse of any element of $G$ is computable by inspection.
\end{proof}

\begin{thrm} \label{th:ramemb-freeness-gen}
Let $G$ be a group with a \ruset.
If \pbratmem{G} is decidable then \pbfree{G} is decidable.
\end{thrm}

\begin{proof}
We show that there exists a Turing reduction from the complementary problem of \pbfree{G} to \pbratmem{G}.
                                                                                        
First, consider a finite subset $X \subseteq G$ with cardinality greater than $1$.
For every $x \in G$, let $A_x$ be the automaton over $G$ defined by:
\begin{itemize}
\item $\ttI$, $\ttQ$, and $\ttT$ are the states of $A_{x}$,
\item the transitions of $A_{x}$ are 
$\transit{\ttI}{x}{\ttQ}$,
$\transit{\ttQ}{\mvg}{\ttT}$,
 and
for each $y \in X$,
$\transit{\ttQ}{y}{\ttQ}$ and
$\transit{\ttT}{y^{-1}}{\ttT}$,
\item $\ttI$ is the unique initial state of $A_x$, and
\item $\ttT$ is the unique terminal state of $A_x$.
\end{itemize}
The behavior of $A_x$ equals $\left\{ x  z  z'^{-1}  : (z, z') \in X^\star \times X^\star \right\}$. 
It thus follows from Lemma~\ref{lem:eq-group} that $X$ is not a code \textiff{} 
there exist $x$, $x' \in X$ such that $x \ne x'$ and  $A_{x}$ accepts $x'$.
If \pbratmem{G} is decidable then $A_x$ is computable from $x$ and $X$ by Lemma~\ref{lem:op-comput}.

Second, consider an element $x \in G$.
Let $B$ be the automaton over $G$ defined by:
\begin{itemize}
\item $\ttI$ and $\ttT $ are the states of $B$,
\item $\transit{\ttI}{x}{\ttT}$ and $\transit{\ttT}{x}{\ttT}$ are the transitions of $B$, 
\item  $\ttI$ is the unique initial state of $B$, and 
\item $\ttT$ is the unique terminal state of $B$.
\end{itemize}
The behavior of $B$ equals $\left\{ x, x^2, x^3, x^4, \dotsc \right\}$.
Therefore, $x$ is torsion \textiff{} $B$ accepts $\mvg$.
Moreover, $B$ is clearly computable from $x$.

The theorem follows from the preceding discussion.
\end{proof}

The proof of Theorem~\ref{th:ramemb-freeness-gen} deserves two observations.
First, the result still holds even if \pbratmem{G} is restricted to those instances $(A, s)$ such that the automaton $A$ has at most $3$ states.
Second, we claim that 
if the inversion in $G$ is computable  then there exists a  many-one reduction from the complementary problem of  \pbfree{G} to \pbratmem{G}; 
the verification is left to the reader.

The following theorem can be proven in the same way as Theorem~\ref{th:ramemb-freeness-gen}.

\begin{thrm} \label{th:ramemb-freeness-pol}
Let $G$ be a group such that (the underlying set of $G$ is recognizable in polynomial time and) the inversion in $G$ is computable in polynomial time. 
If \pbratmem{G} is decidable in polynomial time then \pbfree{G} is decidable in polynomial time.
\end{thrm}

\begin{proof}
Left to the reader.
\end{proof}

The general linear group of degree $d$ over $\Z$ is denoted $\GL{d}{\Z}$:
$$
\GL{d}{\Z} = \left\{ X \in  \carre{\Z}{d} : \det(X) = \pm 1 \right\} \, .
$$
Equivalently, $\GL{d}{\Z}$ is the set of all matrices $X \in  \carre{\Z}{d}$ such that  $X$ has an inverse in $\carre{\Z}{d}$.
Choffrut and Karhum{\"a}ki have shown that \pbratmem{\GL{2}{\Z}} is decidable \cite{ChoffrutK05}.
Hence,  it follows from Theorem~\ref{th:ramemb-freeness-gen}:

\begin{crllr} \label{coro:GL2}
\pbfree{\GL{2}{\Z}} is decidable.
\end{crllr}

Let us now turn to the free group.
To properly deal with this algebraic structure,
we introduce the notion of semi-Thue system.
(Semi-Thue systems are also involved in Section~\ref{sec:semi-Thue-PCP}).

\begin{dfntn}[Semi-Thue system] \label{def:semi-Thue}
A \emph{semi-Thue system} is an ordered pair $T = (\Sigma, R)$,
 where $\Sigma$ is an alphabet and $R$ is a subset of $\Sigma^\star \times \Sigma^\star$.
The elements of $R$ are the \emph{rules} of $T$.
The \emph{immediate accessibility under $T$} is the binary relation over $\Sigma^\star$ defined by:
for every $x$, $y \in \Sigma^\star$,
$y$ is immediately accessible from $x$ under $T$ \textiff{} there exist $(s, t) \in R$ and $z$, $z' \in  \Sigma^\star$ such that $x = z s z'$ and $y = z t z'$.
The reflexive-transitive closure of the immediate accessibility under $T$ is simply called the \emph{accessibility under $T$}.
\end{dfntn}

For the rest of the section, 
overlining is construed as a purely formal operation on symbols.
Let $\Sigma$ be an alphabet and let $\bar \Sigma \defeq \left\{ \bar a : a \in \Sigma \right\}$: 
$\bar \Sigma$ is alphabet such that $\Sigma$ and $\bar \Sigma$ are equinumerous and disjoint.
Given two words $x$ and $y$ over $\Sigmabar$, 
we say that $x$ \emph{freely reduces} to $y$ if $y$ is accessible from $x$ under the semi-Thue system 
$$\left(\Sigmabar, \{ (a \bar a, \mv) : a \in \Sigma  \} \cup  \{ (\bar a a, \mv) : a \in \Sigma  \} \right) \, .$$
A word $w$ over  $\Sigmabar$ is called \emph{freely reduced} if
for every $a \in \Sigma$, neither $a \bar a$ nor $\bar a a$ occurs in $w$
Let $f\colon \left( \Sigmabar \right)^\star \times \left( \Sigmabar \right)^\star \to \left( \Sigmabar \right)^\star$ be defined by:
 for all words $x$ and $y$ over $\Sigmabar$, $f(x, y)$ is the unique freely reduced word over 
$\Sigmabar$ to which $xy$ freely reduces.
The free group over $\Sigma$, denoted $\FG{\Sigma}$, can be defined as follows:
its underlying set is the set of all freely reduced words over $\Sigmabar$
and  
its operation is induced by $f$. 
A more detailed introduction to the free group can be found in \cite{LyndonSchupp}.

Assume now that $\Sigma$ is finite.
Then, the underlying set of $\FG{\Sigma}$ is recognizable in polynomial and the inversion in $\FG{\Sigma}$ is computable in polynomial time.
\pbratmem{\FG{\Sigma}} can be restated as follows:
given a finite  automaton $A$ over $\Sigmabar \cup \{ \mv \}$ and a freely reduced word $s$ over $\Sigmabar$,
decide whether there exists a word $s'$ over $\Sigmabar$ such that $A$ accepts $s'$ and $s'$ freely reduces to $s$.

\begin{dfntn}
Let $\Sigma$ be an alphabet and let $A = (Q, E, I, T)$ be an automaton over $\Sigmabar \cup \{ \mv \}$.

A \emph{free reducibility} of $A$ is an element $(p, q) \in Q \times Q$ for which there exists $a \in \Sigma$ such that the automaton $(Q, E, \{ p \}, \{ q \})$  accepts $a \bar a$ or $\bar a a$.
We say that $A$ is \emph{freely reduced} if for every free reducibility $(p, q)$ of $A$, 
$\transit{p}{\mv}{q}$ belongs to $E$.

Let $\calf$ denote the set of all subsets $F  \subseteq Q \times \{ \mv \} \times Q$ such that $(Q, E \cup F, I, T)$ is freely reduced. 
Note that $\calf$ is non-empty because $Q \times \{ \mv \} \times Q \in \calf$ 
and that $\calf$ is stable under the formation of arbitrary intersections. 
The automaton $\tilde A \defeq (Q,  E \cup \bigcap_{F \in \calf} F, I, T)$ is called the \emph{free reduction} of $A$.
\end{dfntn}

Colloquially, $\tilde A$ is the smallest freely reduced ``super-automaton'' of $A$.

\begin{thrm}[Algorithm~1.3.7 in \cite{Grunschlag99}, see also \cite{Gilman84,Benois69}] \label{th:benois}
Let $\Sigma$ be a finite alphabet.
For every finite automaton $A$ over $\Sigmabar \cup \{ \mv \}$,
\begin{enumerate}[$(i)$.]
 \item $\tilde A$ is computable from $A$ in polynomial time and
 \item the behavior of $\tilde A$ is the closure of the behavior of $A$ under free reduction.
\end{enumerate}
\end{thrm}

Theorem~\ref{th:benois}.$(ii)$ could be stated as follows:
for every word $x$ over $\Sigmabar$, 
$\tilde A$ accepts $x$ 
\textiff{} 
$A$ accepts some word over $\Sigmabar$  that freely reduces to $x$.
From Example~\ref{ex:sigmastar} and Theorem~\ref{th:benois} we deduce:

\begin{crllr} \label{coro:recogn-free}
For any finite alphabet $\Sigma$, 
\pbratmem{\FG{\Sigma}} is decidable in polynomial time.
\end{crllr}

From Theorem~\ref{th:ramemb-freeness-pol} and  Corollary~\ref{coro:recogn-free} we deduce:

\begin{crllr} \label{coro:free-group}
For any finite alphabet $\Sigma$,
\pbfree{\FG{\Sigma}} is decidable in polynomial time.
\end{crllr}

\section{Number of generators}
\label{sec:nbGene}

The section begins with two natural questions:

\begin{open}
Does there exist a semigroup $S_\infty$  with a \ruset{} and  satisfying the following two properties:
\pbfree{S_\infty} is undecidable
and 
\pbkfree{k}{S_\infty} is decidable for every integer $k \ge 1$?
\end{open}

\begin{open}
Let $K$ denote the set of all integers $k \ge 1$ such that there exists a semigroup $S_k$  with a \ruset{} and satisfying the following two properties:  
\pbkfree{k}{S_k} is decidable  and \pbkfree{k + 1}{S_k} is undecidable.
Is the cardinality of $K$ finite? 
\end{open}

Combining Example~\ref{ex:mat-carre-free} above and Corollary~\ref{cor:dim-freeness} below,
we get that $1 \in K$: 
$\carre{\N}{36}$ is a suitable choice for $S_1$.
Combining Example~\ref{ex:zeon-d} above and Theorem~\ref{cor:zeonzeon} below, 
we get that $K \cap \seg{2}{12} \ne \emptyset$: 
for some $k \in \seg{2}{12}$, $\FreeM \times \FreeM$ is a suitable choice for $S_k$.

The following theorem states the existence of bizarre (semi)groups: 

\begin{thrm} \label{th:pathos}
There exists an abelian group $G$ with a (\ruset{} and a) computable operation such that \pbkfree{1}{G} is undecidable.
\end{thrm}

\begin{proof}
 Let $M$ be a universal Turing machine \cite{HopcroftMU01}; 
the input alphabet of $M$ equals $\zeon$. 
 Let $f\colon \FreeM \to \N \cup \{ \infty \}$ be defined by:
 for each $w \in \FreeM$, $f(w)$ equals the running time of $M$ on input $w$.
Note that $f(w)$ is non-zero for every $w \in \FreeM$: 
any Turing machine that decides a non-trivial language must read at least one letter of each input word before halting. 
The following problem is decidable:
given $w \in \FreeM$ and $n \in \N$, decide whether $f(w) \ge n$.

Let $G$ be the set of all $g\colon \FreeM \to \Z$ such that 
$\left\{ w \in \FreeM : g(w) \ne 0 \right\}$ is finite 
and $- f(w) < g(w) < f(w)$ for every $w \in \FreeM$.
Remark that $G$ is a recursive set.
Let us equip $G$ with the computable abelian group operation $\oplus$ defined by:
$$
(g \oplus h)(w) \defeq 
\begin{cases}
 g(w) + h(w) - 2f(w) + 1 & \text{if $g(w) + h(w) \ge  f(w)$} \\
g(w) + h(w) + 2f(w) - 1 & \text{if $g(w) + h(w) \le - f(w)$} \\
g(w) + h(w) & \text{otherwise}
\end{cases} 
$$
for every $g$, $h \in G$ and every $w \in \FreeM$.

It remains to prove that  \pbkfree{1}{G} is undecidable.
Let $w \in \FreeM$.
Let $g \in G$ be defined by: 
$g(w) \defeq 1$ and $g(v) \defeq 0$ for every $v \in \FreeM \setminus \{ w \}$.
Clearly, $g$ generates a subgroup of $G$ with cardinality $2 f(w) - 1$.
Therefore, the following three assertions are equivalent:
$g$ is torsion, $f(w) \ne \infty$, and $M$ halts on input $w$.
Hence, there exists a many-one reduction from the halting problem to \pbkfree{1}{G}.
\end{proof}

For any  commutative  semigroup $S$ with a \ruset{}  and any integer $k \ge 2$, 
\pbkfree{k}{S} is trivially decidable.
Hence, by Theorem~\ref{th:pathos}, there exists a group $G$ with a \ruset{} satisfying the following two properties:
\pbkfree{1}{G} is undecidable and \pbkfree{k}{G} is decidable for every integer $k \ge 2$.

\begin{open} \label{open:pathos-nbgene}
Does there exist a semigroup $S$ with a \ruset{} and an integer $k \ge 2$ satisfying the following two properties: 
\pbkfree{k}{S} is undecidable
and 
\pbkfree{k + 1}{S} is decidable?
\end{open}

\subsection{Regular behaviors}

Theorem~\ref{th:pathos} identifies a ``misbehavior'' of the freeness problems.
The next two propositions ensure that a large class of problems related to the combinatorics of semigroups are well-behaved.

For any set $S$,
 let $\calp(S)$ denote the \emph{power set} of $S$.

\begin{prpstn} \label{prop:monotone}
Let $S$ be a semigroup with a \ruset{} and let $\caly$ be a subset of $\calp(S)$.
For every integer $k \ge 1$, 
let $\rmD(k)$ denote the following problem:
given a $k$-element subset $X \subseteq S$, decide whether $X^+  \in \caly$.
Let $\rmF$ denote the following problem: 
given a finite subset $X \subseteq S$,
decide whether $X \in \caly$.
Assume that the operation of $S$ is computable and that $\rmF$ is decidable.
One of the following two assertions holds:
\begin{enumerate}
 \item For every integer $k \ge 1$, $\rmD(k)$ is decidable.
\item There exists an integer $l \ge 1$ such that 
\begin{itemize}
 \item $\rmD(k)$ is decidable for every integer $k$ with $1 \le k < l$ and 
 \item $\rmD(k)$ is undecidable for every integer $k \ge l$.
\end{itemize}
\end{enumerate}
 \end{prpstn}

\begin{proof}
Let $X$ be a $k$-element subset of $S$.
Put $X^2 \defeq \left\{ x x' : (x, x') \in X \times X \right\}$ and remark that $X^2$ is computable from $X$ because the operation of $S$ is computable. 

First, assume that $X^2$ is not a subset of $X$.
Then, there exist $y \in X^2$ such that $y  \notin X$.
Clearly, $X \cup \{ y \}$ is a $(k + 1)$-element subset of $S$.
Moreover, we have $\left( X \cup \{ y \} \right)^+ = X^+$.
Therefore, 
$X$ is a yes-instance of $\rmD(k)$
\textiff{} 
$X \cup \{ y \}$ is a yes-instance of $\rmD(k+1)$.
Second, assume that $X^2$ is a subset of $X$.
Then, we have $X = X^+$.
Therefore,  $X$ is a yes-instance of $\rmD(k)$ \textiff{} $X$ is a yes-instance of $\rmF$.

It follows from the preceding discussion that $\rmD(k + 1)$ is decidable only if $\rmD(k)$ is decidable.
Hence, the desired result holds true.
\end{proof}

Let us show that Proposition~\ref{prop:monotone} applies to 
mortality, 
semigroup finiteness, and 
semigroup boundedness. 
$\rmD(k)$ equals \pbkmort{k}{S} 
in the case where $\caly$ is the set of all subsets $X \subseteq S$ such that the zero element of $S$ belongs to $X$. 
$\rmD(k)$ equals \pbksemfin{k}{S} 
in the case where $\caly$ is the set of all finite subsets of $S$.
 Let $d$ be a positive integer.
$\rmD(k)$ equals \pbkbound{k}{\carre{\Q}{d}} 
in the case where $S = \carre{\Q}{d}$ and  $\caly$ is  the set of all bounded subsets of $\carre{\Q}{d}$,

The following proposition is a straightforward generalization of Proposition~\ref{prop:monotone}:

\begin{prpstn} \label{prop:monotone-param}
Let $S$ be a semigroup with a \ruset, 
let $A$ be a recursive set,
and let $\caly$ be a subset of $\calp(S) \times A$.
For every integer $k \ge 1$, 
let $\rmD(k)$ denote the following problem:
given a $k$-element subset $X \subseteq S$ and an element $a \in A$, decide whether $(X^+, a)  \in \caly$.
Let $\rmF$ denote the following problem: 
given a finite subset $X \subseteq S$ and an element $a \in A$,
decide whether $(X, a) \in \caly$.
Assume that the operation of $S$ is computable and that $\rmF$ is decidable.
The conclusion is the same as in Proposition~\ref{prop:monotone}.
 \end{prpstn}

\begin{proof}
 The proof is left to the reader.
\end{proof}

Proposition~\ref{prop:monotone}  
can be deduced from Proposition~\ref{prop:monotone-param} by considering the case where $A = \{ \tta \}$.
Let us show that Proposition~\ref{prop:monotone-param} applies to semigroup membership problems and to \pbGPCP.
$\rmD(k)$ equals \pbksemmem{k}{S} in the case where $A = S$ and $\caly$ equals the set of all $(X, a) \in \calp(S) \times S$ such that $a \in X$.  
Lastly, consider the case where 
$S = \FreeM \times \FreeM$, 
$A = \FreeM \times \FreeM \times  \FreeM \times \FreeM$, 
and 
 $\caly$ equals the set of all $(X, (s, s', t, t')) \in \calp(S) \times A$
such that $s x s' = t y t'$ for some $(x, y) \in X \cup \{ (\mv, \mv) \}$.
Then,  $\rmD(k)$ is equivalent to \pbkGPCP{k}.

\pbkGPCP{2} is decidable \cite{HalavaHH02} and \pbkGPCP{5} is undecidable \cite{Claus07},
so there exists $l \in \seg{3}{5}$ such that for every integer $k \ge 1$, 
\pbkGPCP{k} is decidable \textiff{} $k < l$.
Let us illustrate the cases of 
semigroup finiteness, 
semigroup boundedness,
mortality, and
semigroup membership problems with results drawn from the literature.
Let $d$ be a fixed positive integer.
\pbsemfin{\carre{\Q}{d}} is decidable \cite{MandelS77,Jacob77}, 
so for every integer $k \ge 1$, \pbksemfin{k}{\carre{\Q}{d}} is decidable.
\pbkbound{1}{\carre{\Q}{d}} is decidable by Proposition~\ref{prop:power-bound}
and 
\pbkbound{2}{\carre{\Q}{47}} is undecidable \cite{BlondelC03},
so for every integer $k \ge 1$, \pbkbound{k}{\carre{\Q}{47}} is decidable \textiff{} $k = 1$.
\pbkmort{1}{\carre{\Z}{d}} is decidable because for every $M \in \carre{\Z}{d}$, 
$\{ M \}$ is a yes-instance of \pbkmort{1}{\carre{\Z}{d}} \textiff{} $M^d$ is a zero matrix.
On the other hand,  \pbkmort{7}{\carre{\Z}{3}}  is undecidable \cite{Claus07}.
Therefore, there exists $l_0 \in \seg{2}{7}$ such that for every integer $k \ge 1$,
\pbkmort{k}{\carre{\Z}{3}}  is decidable \textiff{} $k < l_0$.
\pbksemmem{1}{\carre{\Q}{d}} is decidable by Remark~\ref{rem:semmem1}  and 
\pbksemmem{k}{S} can be seen as a generalization of \pbkmort{k}{S} for every integer $k \ge 1$ and every semigroup $S$ with a \ruset{} and a zero element.
Therefore, there exists $l_1 \in \seg{2}{l_0}$ such that 
\pbksemmem{k}{\carre{\Z}{3}}  is decidable \textiff{} $k < l_1$.

\subsection{The case of the  freeness problem}

Although Proposition~\ref{prop:monotone} does not apply to freeness problems in any obvious way,
the answer to Question~\ref{open:pathos-nbgene} might as well be ``no''.
Such an eventuality is supported by the next theorem, whose proof relies on the following gadget:

\begin{dfntn}\label{def:gadget-nbgene}
Let $S$ be a semigroup.
For every  integer $d \ge 1$,
every element $x \in S$, and 
every subset $Y \subseteq S$
define
$$
C_d(x, Y) \defeq  \{ x^d \} \cup \bigcup_{i = 0}^{d - 1} x^i Y \, .
$$
\end{dfntn}

The following two lemmas establish the main properties of the gadget.

\begin{lmm} \label{lem:Cd-a-sigma}
Let $d$ be a positive integer,
let $a$ be a symbol, and 
let $\Sigma$ be an alphabet such that $a \notin \Sigma$.
\begin{enumerate}[$(i)$.]
\item 
Assume that $\Sigma$ is finite. 
Let $k$ denote the cardinality of $\Sigma$.
The cardinality of $C_d(a, \Sigma)$ equals $kd + 1$.
\item 
The language $C_d(a, \Sigma)$ is a prefix code.
\item 
Every non-empty word over $\{ a \} \cup  \Sigma$  that does not end with $a$ belongs to ${C_d(a, \Sigma)}^+$.
\end{enumerate}
\end{lmm}

\begin{proof}
Parts~$(i)$ and~$(ii)$ are clear. 
Let $(n, b)\in \N \times \Sigma$. 
Write $n$ in the form $n = q d + r$ with $q \in \N$ and $r \in \seg{0}{d - 1}$.
Since $a^d$ and $a^r b$ belong to  $C_d(a,\Sigma)$, 
$a^n b = {(a^d)}^q (a^r b)$ is an element of ${C_d(a, \Sigma)}^+$.
Put $L \defeq \left\{ a^n b : (n, b) \in \N \times \Sigma \right\}$.
We have just proven $L \subseteq {C_d(a, \Sigma)}^+$.
It follows $L^+ \subseteq {C_d(a, \Sigma)}^+$.
Since $L^+$ equals the set of all non-empty words over $\{ a \} \cup  \Sigma$ that do not end with $a$, 
part~$(iii)$ holds.
\end{proof}

In fact,  it is easy to see that 
${C_d(a, \Sigma)}^+ =  {(\{ a \} \cup  \Sigma)}^\star \Sigma {\{ a^d \}}^\star \cup {\{ a^d \}}^+ $.

\begin{lmm} \label{lem:gadget-code}
Let $S$ be a semigroup, 
let $d$ be a positive integer,
let $x$ be an element of $S$, and 
let $Y$ be a finite subset of $S$ such that $x \notin Y$.
Let $k$ denote the cardinality of $Y$.
The set $\{ x \} \cup Y$ is a code \textiff{} the following two conditions are met:
the cardinality of $C_d(x, Y)$ equals $k d + 1$ 
and 
$C_d(x, Y)$ is a code.
\end{lmm} 

\begin{proof}
The ``only if part'' follows from Lemmas~\ref{lem:Cd-a-sigma}.$(i)$ and~\ref{lem:Cd-a-sigma}.$(ii)$:
if $\{ x \} \cup Y$ is a code then $x$ can be thought as the symbol $a$ and $Y$ as the alphabet $\Sigma$.
Let us now prove the ``if part''. 

Let $\Sigma$ be an alphabet with cardinality $k$,
let $a$ be a symbol such that $a \notin \Sigma$, 
and 
let $\sigma\colon {(\{ a \} \cup \Sigma)}^+ \to S$ be a morphism such that $\sigma(a) = x$ and $\sigma(\Sigma) = Y$.
 Clearly, $\sigma$ maps $C_d(a, \Sigma)$ onto $C_d(x, Y)$.
Assume that the cardinality of $C_d(x, Y)$  equals $kd + 1$ and that $C_d(x, Y)$ is a code.
Then, by Lemma~\ref{lem:Cd-a-sigma}.$(i)$, $\sigma$ induces a bijection from $C_d(a, \Sigma)$ onto $C_d(x, Y)$, 
and subsequently, it follows from Claim~\ref{clm:pre-image-code} that $\sigma$ is injective on ${C_d(a, \Sigma)}^+$.
Let $u$, $v \in {(\{ a \} \cup \Sigma)}^+$ be such that $\sigma(u) = \sigma(v)$.
Let $b \in \Sigma$. 
On the one hand, we have $\sigma(ub) = \sigma(vb)$,
and on the other hand, both $ub$ and $vb$ belong to ${C_d(a, \Sigma)}^+$ by Lemma~\ref{lem:Cd-a-sigma}.$(iii)$.
Since $\sigma$ is injective on the latter set, we get $ub = vb$, which implies $u = v$.
We have thus shown that $\sigma$ is injective.
Therefore, $\{ x \} \cup Y$ is a code by Claim~\ref{clm:pre-image-code}.
\end{proof}

\begin{thrm} \label{th:nbGene}
Let $S$ be a semigroup with a computable operation and let $k$ and $d$ be positive integers.
If \pbkfree{kd + 1}{S} is decidable then  \pbkfree{k + 1}{S} is decidable. 
\end{thrm}

 \begin{proof}
For any element $x \in S$ and any $k$-element subset $Y \subseteq S$, 
$C_d(x, Y)$ is computable from  $x$ and $Y$  because the operation of $S$ is computable. 
Hence, it follows from Lemma~\ref{lem:gadget-code} that there exists a many-one reduction from \pbkfree{k + 1}{S} to \pbkfree{kd + 1}{S}.
 \end{proof} 

If  \pbkfree{k_0}{S} is undecidable for some integer $k_0 \ge 2$ then it follows from Theorem~\ref{th:nbGene} that  \pbkfree{1 + (k_0 - 1)d}{S} is undecidable  for every integer $d \ge 1$.

\begin{crllr} \label{coro:nb-gene}
Let $S$ be a semigroup with a computable operation.
\begin{enumerate}[$(i)$.]
\item 
If there exists an integer $k \ge 2$ such that \pbkfree{k}{S} is decidable then \pbkfree{2}{S} is decidable.
\item 
If there exists an odd integer $k \ge 3$ such that \pbkfree{k}{S} is decidable then \pbkfree{3}{S} is decidable.
\end{enumerate} 
\end{crllr}

  \section{Two-by-two matrices} \label{sec:dec-two-two}
  
The most exciting open questions about the decidability of freeness problems arise from two-by-two matrix semigroups \cite{GawrychowskiGK10,BlondelCK04,CassaigneHK99,Klarner91}.

It is noteworthy that matrix mortality is also tricky in dimension two.
In 1970, 
Paterson introduced \pbmort{\carre{\Z}{3}} and showed that the problem  is undecidable \cite{Paterson70}.
Since then, 
the decidability of \pbmort{\carre{\Z}{2}} has been repeatedly reported as an open question \cite{BB02,HalavaH01,Claus07,Miller94,KromKrom90,Schultz77}.
The only partial results obtained so far are:
\pbkmort{2}{\carre{\Z}{2}} is decidable \cite{BB02}
and
\pbmort{\carre{\N}{d}} is decidable for each integer $d \ge 1$ \cite{BlondelT97}.

\subsection{Toward undecidability} \label{sec:to-undec-two-two}

Although the decidabilities of \pbfree{\carre{\N}{2}},  \pbfree{\carre{\Z}{2}}, and  \pbfree{\carre{\Q}{2}} are still open, 
Bell and Potapov have proven that \pbkfree{7}{\carre{{\Hamilton}}{2}}  is undecidable, 
where  
$$
\Hamilton \defeq 
\left\{ 
 \begin{bmatrix}
 x  &  y &  z &  t \\
-y &  x &  -t & z \\
-z & t &  x &  -y \\
-t &  -z & y &  x
 \end{bmatrix} 
: x, y, z, t \in \Q 
\right\} 
$$
is the skew field of rational quaternions  \cite{BellP08}.
Besides, it follows from Theorem~\ref{th:MandelSimon} that  \pbkfree{1}{\carre{{\Hamilton}}{2}} is decidable: 
for every $M \in \carre{\Hamilton}{2}$, $M$ is torsion \textiff{} $M^{8} = M^{8 + r(8)}$.
A natural question is thus:

\begin{open} \label{open:Q-2-2}
Does there exist a commutative semiring  $D$ with a \ruset{} and 
satisfying the following two properties:
  \pbkfree{1}{\carre{D}{2}} is decidable
and
\pbfree{\carre{D}{2}} is undecidable?
\end{open}

 Let $D$ be a semiring with a \ruset{} such that \pbkfree{1}{\carre{D}{2}} is decidable.
 Then, 
 the set of those elements of $D$ that are torsion under multiplication is recursive:
 for every $t \in D$, 
 $t$ is torsion under multiplication 
 \textiff{}
 $\smalltwotwo{t}{0}{0}{1}$ is torsion under matrix multiplication.
 Moreover, the set of those elements of $D$ that are torsion under addition is also recursive:
 for every $t \in D$, 
 $t$ is torsion under addition 
 \textiff{}
 $\smalltwotwo{1}{t}{0}{1}$ is torsion under matrix multiplication.
 Hence, the decidability of \pbkfree{1}{\carre{D}{2}} nicely polices $D$.

Let $K$ be an extension field of $\Q$ with degree $d$.
Since there exists an injective ring homomorphism from $K$ to $\carre{\Q}{d}$ \cite{Janusz73}, 
Theorem~\ref{th:MandelSimon} ensures that for every $M \in \carre{K}{2}$, $M$ is torsion \textiff{} $M^{2d} = M^{2d + r(2d)}$.
Therefore, \pbkfree{1}{\carre{K}{2}} is decidable.
Proving the undecidability of \pbfree{\carre{K}{2}} for some field extension $K$ of $\Q$ would solve Question~\ref{open:Q-2-2} and be a significant advance towards proving the undecidability of \pbfree{\carre{\Q}{2}}.

Let us now introduce a more general question than Question~\ref{open:Q-2-2}.

\begin{lmm}  \label{lem:det-nonzero-twotwo}
Let $A$ be a commutative ring and let $X \in \carre{A}{2}$ be such that the determinant of $X$  equals $0$.
\begin{enumerate}[$(i)$.]
 \item For every $Y \in \carre{A}{2}$, equality $XXYX = XYXX$ holds.
  \item The matrix $X$ is torsion \textiff{} its trace is torsion under multiplication.
 \end{enumerate}
\end{lmm}

\begin{proof}
Let $t$ denote the trace of $X$.

The characteristic polynomial of $X$ equals 
$\ttz^2 - t \ttz$,
so $X^2 = t X$ by the Cayley-Hamilton theorem.
It follows that $ X X Y X =  t X Y X = X Y X X$ for every $Y \in \carre{A}{2}$, and thus part~$(i)$ holds.

Let us now turn to part~$(ii)$.
On the one hand, we have $X^{n + 1} = t^n X$ for every $n \in \N$.
Therefore, $t$ is torsion only if $X$ is torsion.
On the other hand, the trace of $X^n$ equals $t^n$ for every integer $n \ge 1$.
Therefore,  $X$ is torsion only if $t$ is torsion.
We have thus shown part~$(ii)$.
\end{proof}

Part~$(i)$ of Lemma~\ref{lem:det-nonzero-twotwo}  previously appeared in \cite{CassaigneHK99}.

Let $K$ be a field.
The \emph{general linear group} of degree $d$ over $K$ is denoted $\GL{d}{K}$: 
$$
\GL{d}{K} = \left\{ X \in \carre{K}{d} : \det(X) \ne 0 \right\} \, .
$$
Assume that the underlying set of $K$ is recursive and that the addition and the multiplication of $K$ are computable.
By Lemma~\ref{lem:op-comput}.$(ii)$, the additive inversion in $K$ and the multiplicative inversion in $K \setminus \{ 0 \}$ are also computable. 
In particular, determinants of matrices over $K$ are computable, and thus $\GL{d}{K}$ is a recursive set.

\begin{prpstn} \label{prop:twotwo-Q-inversible}
Let $K$ be a field with computable operations.
\pbfree{\carre{K}{2}} is decidable \textiff{}  \pbfree{\GL{2}{K}} is decidable
\end{prpstn} 

\begin{proof}
The ``only if part'' is trivial since $\GL{2}{K}$ is a subsemigroup of $\carre{K}{2}$.
Let us now prove the ``if part''.

Assume that \pbfree{\GL{2}{K}} is decidable.
Given a finite subset $\calx \subseteq \carre{K}{2}$, 
let us explain how to decide whether $\calx$ is a code. 
The case where $\calx$ is a subset of $\GL{2}{K}$ is trivial.
If $\calx$ is not a subset of $\GL{2}{K}$ and if the cardinality of $\calx$ is greater than $1$ then $\calx$ is not a code by Lemma~\ref{lem:det-nonzero-twotwo}.$(i)$.
It remains to deal with the case where $\calx = \{ X \}$ for some $X \notin \GL{2}{K}$.
We rely on Lemma~\ref{lem:det-nonzero-twotwo}.$(ii)$.
Let $t$ denote the trace of $X$.
If $t = 0$ then $X$ is torsion (and even nilpotent).
If $t \ne 0$ then $\smalltwotwo{t}{0}{0}{1}$ belongs to $\GL{2}{K}$, 
and moreover, 
$X$ is torsion
\textiff{}
$\smalltwotwo{t}{0}{0}{1}$ is torsion.
\end{proof}

Note that the decidability of \pbfree{\GL{2}{\Q}} is \emph{not}  trivially implied by Corollary~\ref{coro:GL2}: 
the structure of $\GL{2}{\Q}$ is far more complicated than the one of $\GL{2}{\Z}$.

\begin{crllr} \label{coro:accept-free}
Let $K$ be a field with computable operations.
If \pbfree{\carre{K}{2}} is undecidable then  \pbratmem{\carre{K}{2}} is undecidable.
\end{crllr} 

\begin{proof}
Assume that  \pbratmem{\carre{K}{2}} is decidable.
Then, its restriction \pbratmem{\GL{2}{K}} is decidable.
Since $\GL{2}{K}$ is a group, it follows from Theorem~\ref{th:ramemb-freeness-gen} that  \pbfree{\GL{2}{K}}  is decidable.
Thus, Proposition~\ref{prop:twotwo-Q-inversible} ensures that \pbfree{\carre{K}{2}} is decidable.
\end{proof}

The reader who conjectures that the answer to Question~\ref{open:Q-2-2} is ``yes'' might want to first tackle:


\begin{open} \label{open:accept}
Does there exist a commutative semiring $D$ with a \ruset{} and satisfying the following two properties: \pbkfree{1}{\carre{D}{2}} is decidable and \pbratmem{\carre{D}{2}} is undecidable?
\end{open}

\subsection{Toward decidability} 
\label{sec:to-dec-two-two}

This section focuses on the following open question.

\begin{open}[\cite{BlondelCK04,CassaigneHK99}] \label{open:k-two-two}
\label{open:twotwo}
Is \pbkfree{2}{\carre{\N}{2}}  decidable?
\end{open}

Note that if \pbkfree{k}{\carre{\N}{2}} is decidable for some integer $k \ge2$ then, by Corollary~\ref{coro:nb-gene}.$(i)$, 
\pbkfree{2}{\carre{\N}{2}} is decidable.

\subsubsection{Two upper-triangular matrices}
\label{sec:two-up}

For each semiring $D$ and each integer $d \ge 1$,
 let $\Tri(d, D)$ denote the  set of all $d$-by-$d$ upper-triangular matrices over $D$:
$\Tri(d, D)$ is a subsemiring of $\carre{D}{d}$, 
so in particular, $\Tri(d, D)$ is a multiplicative semigroup.
For instance, $\Tri(2, D)$ is the set of all matrices of the form
$\smalltwotwo{a}{b}{0}{c}$ with $a$, $b$, $c \in D$.

\begin{open} \label{open:twotwo-triangle}
Is  \pbkfree{2}{\Tri(2, \N)} decidable?
\end{open}

For all integers $k$, $d \ge 1$, 
\pbkfree{k}{\Tri(d, \Q)} is decidable \textiff{} 
\pbkfree{k}{\Tri(d, \Z)} is decidable:
the proof is the same as for Theorem~\ref{th:Z-Q}.
In particular,  \pbkfree{2}{\Tri(2, \Q)} is decidable \textiff{} \pbkfree{2}{\Tri(2, \Z)}  is decidable.
Put 
\begin{align*}
D_\lambda & \defeq \twotwo{\lambda}{0}{0}{1}
& &  \text{and}  &  
T_\lambda & \defeq  \twotwo {\lambda}{1}{0}{1} \, .
\end{align*} 
for each $\lambda \in \C$.

\begin{xmpl} \label{ex:2-by-2-codes}
The sets 
$\left\{ D_2, T_2 \right\}$,
$\left\{ D_2, T_3 \right\}$, and 
$\left\{ D_{2 \sur 7}, T_{3 \sur 4} \right\}$
are codes under matrix multiplication \cite{CassaigneHK99}.
\end{xmpl} 

\begin{xmpl}
The sets 
$\left\{ D_2, T_{1 \sur 2} \right\}$ 
and  
$\left\{ D_{2 \sur 3}, T_{-3 \sur 5} \right\}$
are not codes under matrix multiplication since 
$
D_2 T_{1 \sur 2}
= \smalltwotwo{1}{2}{0}{1}
=  T_{1 \sur 2} D_2  T_{1 \sur 2} D_2
$
\cite{CassaigneHK99}
and 
$D_{2 \sur 3} T_{-3 \sur 5} D_{2 \sur 3} T_{-3 \sur 5} = 
 \smalltwotwo{4 \sur 25}{2 \sur 5}{0}{1} =
T_{-3 \sur 5}  T_{-3 \sur 5} D_{2 \sur 3} D_{2 \sur 3}$
\cite{GawrychowskiGK10}.
\end{xmpl}

Let $\Pi$ denote the set of all $(\lambda, \mu) \in \C \times \C$ such that  
$\{ D_\lambda, T_\mu \}$ is not a code under matrix multiplication.
 One reason why it might be easier to deal with triangular matrices is that
 \pbkfree{2}{\Tri(2, \Q)} reduces to recognizing $\Pi \cap (\Q \times \Q)$ \cite{CassaigneHK99}.
Moreover, for all $\lambda$, $\mu \in \C \setminus \{ 0, 1 \}$,
the following four assertions are equivalent:
$(\lambda, \mu) \in \Pi$,
$(\mu, \lambda) \in \Pi$,
$(\lambda^{-1}, \mu^{-1})\in \Pi$,
and 
$(\mu^{-1}, \lambda^{-1} ) \in \Pi$ \cite{CassaigneHK99}.

Two partial algorithms for recognizing $\Pi \cap (\Q \times \Q)$ have been proposed \cite{CassaigneHK99,GawrychowskiGK10}.
The latest one, which is by Gawrychowski, Gutan, and Kisielewicz \cite{GawrychowskiGK10},
seems more efficient in practice:
it solves the following example much faster than the older algorithm.

\begin{xmpl} \label{ex:pb-open}
Put $D \defeq D_{2 \sur 3}$ and $T \defeq  T_{3 \sur 5}$.
The set $\{ D, T  \}$ is not a code under matrix multiplication because  both products
$$
DTTTTTTTTTTDDTDDTDDDDDDDDDD 
$$
and 
$$
TTDDDDDDTTDDTDTDTDDTTDDTDTT
$$
are equal to
$$
\twotwo{\frac{32768}{6591796875}}{\frac{242996824}{146484375}}{0}{1} \, .
$$
Note that $D$ and $T$  satisfy no shorter non-trivial equation \cite{GawrychowskiGK10}.
\end{xmpl}

In addition to showing a surprising combinatorial explosion,
Example~\ref{ex:pb-open} answers an open question from \cite{BlondelCK04,CassaigneHK99}.


\subsubsection{One upper-triangular and one lower-triangular matrix}

\label{sec:LucGuyot}

Put 
\begin{align*}
A_{\lambda} & \defeq \twotwo{1}{\lambda}{0}{1}
& &  \text{and}  &  
B_{\lambda}  & \defeq \twotwo {1}{0}{\lambda}{1} 
\end{align*}
for each $\lambda \in \C$.
Let $\Lambda$ denote the set of all $\lambda \in \C$ such that $\left\{ A_\lambda, B_\lambda  \right\}$ is not a code under matrix multiplication. 
The study of $\Lambda$ was initiated by Brenner and Charnow \cite{BrennerC78}.
Our motivation to continue is that \pbkfree{2}{\carre{\Q}{2}} is decidable only if $\Lambda \cap \Q$ is recursive.
We first prove that $\Lambda = - \Lambda$.

\begin{lmm} \label{lem:Lambda-sym}
 For every $\lambda \in \C$,
$\left\{ A_\lambda, B_\lambda  \right\}$ is a code  under matrix multiplication
 \textiff{}  
$\left\{ A_{- \lambda}, B_{- \lambda} \right\}$ is a code under matrix multiplication.
\end{lmm}

\begin{proof}
For every group $G$  and every subset $X \subseteq G$, 
$X$ is a code \textiff{} $\left\{ x^{-1} : x \in X \right\}$ is a code.
Since $A_\lambda^{-1} = A_{-\lambda}$ and $B_\lambda^{-1} = B_{-\lambda}$ for every $\lambda \in \C$,
the desired result holds.
\end{proof}

Let us now prove that every element of $\Lambda \cap \R$ is comprised between $-1$ and $+1$ exclusive.

\begin{prpstn} \label{prop:mobius-code}
For every real number $\lambda$ with $| \lambda | \ge 1$, 
$\left\{ A_\lambda, B_\lambda  \right\}$
is a code under matrix multiplication.
\end{prpstn}

\begin{proof}
By Lemma~\ref{lem:Lambda-sym}, we only have to prove that $\left\{ A_\lambda, B_\lambda  \right\}$ is a code
for every real number $\lambda \ge 1$.

Let $\mathcal{A}$ denote the set of all $\smalltwoone{x}{y} \in \R^{2 \times 1}$ such that $0 < y < x$.
Let $\mathcal{B}$ denote the set of all $\smalltwoone{x}{y} \in \R^{2 \times 1}$ such that $0 < x < y$.
Remark that for all real numbers $x$, $y > 0$,
$A_\lambda \smalltwoone{x}{y}$ belongs to $\mathcal{A}$ 
while 
$B_\lambda \smalltwoone{x}{y}$ belongs to $\mathcal{B}$. 
Let $M$, $N \in \left\{ A_\lambda, B_\lambda  \right\}^\star$.
From the previous remark, we deduce that  
 $A_\lambda M\smalltwoone{1}{1}$ belongs to $\mathcal{A}$
while 
$B_\lambda N \smalltwoone{1}{1}$ belongs to $\mathcal{B}$.
Since $\mathcal{A} \cap \mathcal{B} = \emptyset$, 
we have $A_\lambda M \ne  B_\lambda N$.
The desired result now follows from Lemma~\ref{lem:eq-group}.
\end{proof}

Note that $\Lambda$ contains complex numbers with moduli $1$ or more.

\begin{xmpl}
Let $i$ denote the imaginary unit.
The sets 
$\left\{ A_i, B_i \right\}$ 
and 
$\left\{ A_{3i \sur 2}, B_{3i \sur 2}  \right\}$ 
are not codes under matrix multiplication because 
$$
A_iB_iA_i = \twotwo{0}{i}{i}{0} = B_i A_i B_i \, .
$$ 
and 
$$
AABB AB BABA  
=  
\twotwo{-41 \sur 4}{33i \sur 8}{33i \sur 8}{25 \sur 16}
=
BABA AB AABB  \,,
$$
with $A \defeq A_{3i \sur 2}$ and $B \defeq B_{3i \sur 2}$.
\end{xmpl}

Finally, let us prove the main result of the section: 
the supremum of $\Lambda \cap \Q$ equals $1$.

\begin{lmm}[Brenner and Charnow \cite{BrennerC78}] \label{lem:lambda-sqrt}
Let $\lambda$ be a real number.
If there exist two integers $m$, $n \ge 1$ such that 
\begin{equation} \label{eq:lambda-m-n-racine}
\lambda^2 =    \frac{ mn  - m  - n - 1}{m  n} 
\end{equation}
then   $\left\{ A_{\lambda}, B_\lambda  \right\}$ is not a code under matrix multiplication.
\end{lmm}

\begin{proof}
Let $m$, $n \in \Z$.
It is easy to check that 
$A_{\lambda}^m = A_{m \lambda}$, 
$B_{\lambda}^n = B_{n \lambda}$,
$$
A_\lambda B_\lambda^n A_\lambda^m B_\lambda   = 
\twotwo{mn \lambda^4 + (m + n + 1)  \lambda^2 + 1} {mn \lambda^3 + (m + 1) \lambda}
{mn \lambda^3 + (n + 1) \lambda} {m n \lambda^2 + 1} \, ,
$$
and 
$$
B_\lambda A_\lambda^m B_\lambda^n A_\lambda  = 
\twotwo{mn \lambda^2 + 1}{mn \lambda^3 + (m + 1) \lambda}
{mn \lambda^3 + (n + 1) \lambda}{mn \lambda^4 + (m + n + 1) \lambda^2 + 1} \, .
$$
It follows that $A_\lambda B_\lambda^n A_\lambda^m B_\lambda = B_\lambda A_\lambda^m B_\lambda^n A_\lambda$ \textiff{} $m n \lambda^2 + 1 = mn \lambda^4 + (m + n + 1) \lambda^2 + 1$.
Therefore,  $B_\lambda A_\lambda^m B_\lambda^n A_\lambda = A_\lambda B_\lambda^n A_\lambda^m B_\lambda$ holds whenever $m$ and $n$ satisfy Equation~\eqref{eq:lambda-m-n-racine}. 
\end{proof} 

For each integer $n \ge 3$, 
put $\lambda_n \defeq   \sqrt{1 - 2n^{-1} - n^{-2}}$.
On the one hand, 
$\lambda_n$ tends to $1$ as $n$ tends to infinity.
On the other hand,  
it follows from Lemma~\ref{lem:lambda-sqrt} that $\lambda_n \in \Lambda$: 
consider the special case where $m = n$ in Equation~\eqref{eq:lambda-m-n-racine}. 
We have thus proven that the supremum of $\Lambda \cap \R$  equals $1$.
However, $\lambda_n$ is irrational.

\begin{prpstn}
For every real number $\delta > 0$, 
there exists $\lambda \in  \Q$ such that  $1 - \delta < \lambda < 1$ and $\{  A_\lambda, B_\lambda \}$ is not a code under matrix multiplication.
\end{prpstn}

\begin{proof}
Let $\left(  n_0, n_1, n_2, n_3, \dotsc \right)$  be  the sequence of integers inductively defined by:
$n_0 = 3$, $n_1 =  6$ and $n_{k + 2} = 6 n_{k + 1} - n_k - 6$ for every $k \in \N$.
It is easy to check that:
$$
n_k = \frac{3}{4} \left( {( 3 + 2 \sqrt{2} )}^ k + {(3 - 2 \sqrt{2})}^ k  \right) + \frac{3}{2}
$$
for every $k \in \N$.
Hence, $n_k$ is positive for every $k \in \N$ and   
$$
\lambda_k \defeq  1 - \frac{n_{k + 1} + n_k + 3}{2 n_{k + 1} n_k}
$$
is a rational number that tends to $1$ as $k$ tends to infinity.
Now, remark that the bivariate polynomial
$$
p( \ttx, \tty) \defeq \ttx^2 + \tty^2 -6 \ttx \tty  + 6 \ttx + 6\tty + 9 
$$
satisfies:
\begin{equation}
 p(6\ttx- \tty -6, \ttx) = p(\ttx, \tty)     \label{eq:p-sym-trans} 
\end{equation} 
and
\begin{equation}
\left( 1 - \frac{\ttx + \tty + 3}{2 \ttx \tty} \right)^2
- 
\frac{ \ttx \tty - \ttx - \tty - 1}{\ttx \tty}  
= \frac{ p(\ttx, \tty) }{ 4 \ttx^2 \tty^2 } \label{eq:p-diophante}
 \, .
\end{equation} 
Relying on Equation~\eqref{eq:p-sym-trans}, it is easy to check by induction that  $p(n_{k + 1}, n_k) = 0$ for every $k \in \N$.
Therefore, Equation~\eqref{eq:p-diophante} ensures that 
$$
\lambda_k ^2 = \frac{ n_{k + 1} n_k - n_{k + 1} - n_k - 1}{n_{k + 1} n_k}  \, , 
$$
and thus  $\{  A_{\lambda_k}, B_{\lambda_k} \}$ is not a code by Lemma~\ref{lem:lambda-sqrt}.
\end{proof}

Let $\Lambda'$ denote the set of all $\lambda \in \C$ such that $\left\{ A_\lambda, B_\lambda, A_{- \lambda}, B_{- \lambda}  \right\}^\star$ is not a free group.
A large literature is devoted to the study of $\Lambda'$.
It is clear 
that $\Lambda \subseteq \Lambda'$, 
that $- \Lambda' = \Lambda'$, and 
that no transcendental number belongs to $\Lambda'$.
Moreover, 
it is well-known that for every $\lambda \in \Lambda'$, $| \lambda | < 2$ \cite{LyndonSchupp}.
Many rational and algebraic numbers have been identified in $\Lambda'$ \cite{Beardon93,Grytczuk00}: 
in particular, the supremum of $\Lambda' \cap \R$ equals $2$  \cite{Beardon93}.
However, the existence of a rational number $\lambda$  satisfying $0 < | \lambda | < 2$ and $\lambda \notin \Lambda'$ is a long-standing open question \cite{LyndonSchupp,Grytczuk00}.
Similarly, we state:

\begin{open}[Guyot \cite{GuyotPriv08}] \label{open:luc-guyot}
Is there any rational number $\lambda$ with $| \lambda | < 1$  such that $\left\{ A_\lambda, B_\lambda  \right\}$
is a code under matrix multiplication?
\end{open}

\subsection{Substitutions over the binary alphabet} 
\label{sec:two-two-morph}

In this section, we  examine:

\begin{open} \label{open:k-morph-two}
Is \pbkfree{2}{\hom(\FreeM)} decidable?
\end{open}

(The notation $\hom$ is introduced in Definition~\ref{def:hom-Sigma}.)
To motivate the introduction of Question~\ref{open:k-morph-two}, 
let us consider the function from $\hom(\FreeM)$ to $\carre{\N}{2}$ that maps each $\sigma \in \hom(\FreeM)$ to 
 $$ 
 P_\sigma \defeq 
  \twotwo
  {\nocc{\ze}{\sigma(\ze)}}{\nocc{\ze}{\sigma(\on)}}
  {\nocc{\on}{\sigma(\ze)}}{\nocc{\on}{\sigma(\on)}}  
 $$
(according to Definition~\ref{def:inc-mat}, $P_\sigma$ is the incidence matrix of  $\sigma$ relative to $\ze \on$).
The considered function is clearly surjective, 
it is a morphism by Claim~\ref{claim:mat-morph}.$(i)$,
and 
it is ``almost injective'' by Claim~\ref{claim:mat-morph}.$(ii)$.
Therefore, the semigroups $\hom(\FreeM)$ and $\carre{\N}{2}$ have very similar structures, 
and thus Questions~\ref{open:k-two-two} and \ref{open:k-morph-two} are likely similar.
However, we do not know whether Question~\ref{open:k-morph-two} is easier or harder to solve than Question~\ref{open:k-two-two}.

The following claim is an immediate corollary of Claim~\ref{claim:mat-morph}.$(i)$;
it provides a simple way to generate yes-instances of \pbkfree{2}{\hom(\FreeM)} from yes-instances of \pbkfree{2}{\carre{\N}{2}}.

\begin{clm} \label{prop:mat-morph}
Let $\sigma$, $\tau \in \hom(\FreeM)$ be such that $P_{\sigma} \ne P_{\tau}$.
If $\left\{ P_{\sigma}, P_{\tau} \right\}$ is a code under matrix multiplication 
then $\{ \sigma, \tau \}$ is a code under function composition.
\end{clm} 

For instance, let us construct four yes-intances of \pbkfree{2}{\hom(\FreeM)} from Example~\ref{ex:2-by-2-codes}:

\begin{xmpl}
For each $p \in \N$, let $\delta_p$, $\tau_p$, $\tau'_p \in \hom(\FreeM)$ be defined by:
 \begin{align*}
 & 
 \begin{cases}
 \delta_p(\ze)  \defeq  \ze^p \\
 \delta_p(\on)  \defeq  \on 
 \end{cases} ,
 & 
 &  
 \begin{cases}
 \tau_p(\ze) \defeq \ze^p \\
 \tau_p(\on) \defeq \on \ze
 \end{cases} ,
 & &
\text{and}
 &  
 \begin{cases}
 \tau'_p(\ze) \defeq \ze^p \\
 \tau'_p(\on) \defeq \ze \on 
 \end{cases} 
 .
 \end{align*} 
In the notation of Section~\ref{sec:two-up}, 
we have $P_{\delta_p} = D_p$ and $P_{\tau_p} = P_{\tau'_p} = T_p$ for every $p \in \N$.
It then follows from Example~\ref{ex:2-by-2-codes} and Claim~\ref{prop:mat-morph} that 
$\{ \delta_2, \tau_2 \}$, 
$\{ \delta_2, \tau_3 \}$, 
$\{ \delta_2, \tau'_2 \}$, and 
$\{ \delta_2, \tau'_3 \}$ 
are codes under function composition.
\end{xmpl}

The next two yes-instances of \pbkfree{2}{\hom(\FreeM)} cannot be obtained by applying the previous method.
Note that testing the injectivity of a given morphism $\sigma \in \hom(\FreeM)$ is trivial: 
$\sigma$ is injective \textiff{} $\sigma(\ze\on) \ne \sigma(\on \ze)$ (see Example~\ref{ex:zeon-d}).
Recall also that 
injective functions are left-cancellative under composition (Example~\ref{ex:simplifiable})
and that 
left-cancellability occurs in the hypotheses of Lemma~\ref{lem:eq-group}.

\begin{xmpl}
 Let  $\upsilon$, $\upsilon' \in \hom(\FreeM)$ be defined by:
\begin{align*}
& 
\begin{cases}
\upsilon(\ze) \defeq  \ze \on  \\
\upsilon(\on) \defeq \ze \on \on 
\end{cases}
& &
\text{and}
& 
\begin{cases}
\upsilon'(\ze) \defeq \on \ze \\
\upsilon'(\on) \defeq \on \on \ze 
\end{cases}
.
\end{align*}
For any $x \in \zeon^+$, $\upsilon(x)$ begins with $\ze$ while $\upsilon'(x)$ begins with $\on$.
Therefore, for any $\alpha$, $\alpha' \in \hom(\FreeM)$,
we have 
$\upsilon \alpha \ne  \upsilon' \alpha'$
unless 
$\alpha(\ze) = \alpha(\on) = \alpha'(\ze) = \alpha'(\on) = \mv$.
It then follows from Lemma~\ref{lem:eq-group} that $\{ \upsilon, \upsilon' \}$ is a code under function composition.
However, remark that $P_{\upsilon} = P_{\upsilon'} = \smalltwotwo{1}{1}{1}{2}$.
\end{xmpl}

\begin{xmpl}
 Let  $\phi$, $\mu  \in \hom(\FreeM)$ be defined by:
\begin{align*}
 & 
 \begin{cases}
 \phi(\ze)  \defeq  \ze \on \\
 \phi(\on)  \defeq  \ze 
 \end{cases} 
& &
\text{and}
& 
 \begin{cases}
 \mu(\ze) \defeq \ze \on \\
 \mu(\on) \defeq \on \ze
 \end{cases} 
 & 
.
\end{align*}
Morphisms $\phi$ and $\mu$ play a central role in combinatorics of words \cite{Lothaire1,Lothaire2};
they are usually called the \emph{Fibonacci substitution} and the \emph{Thue-Morse substitution}, respectively. 
Let $\alpha$, $\beta \in {\{ \phi, \mu \}}^+$.
It is easy to see that $\alpha(\ze)$ and $\beta(\ze)$ begin with $\ze \on$.
Therefore, 
$(\phi \alpha)(\ze)$ begins with $\ze \on \ze$ 
while 
$(\mu \beta)(\ze)$ begins with $\ze \on \on$.
It follows that $\phi \alpha \ne \mu \beta$.
Hence, Lemma~\ref{lem:eq-group} implies that $\{ \phi, \mu \}$ is a code under function composition.
However, remark that 
$P_\phi = \smalltwotwo{1}{1}{1}{0} \ne \smalltwotwo{1}{1}{1}{1} = P_\mu$
and 
$P_ \mu P_\mu P_\phi P_\mu = P_ \mu P_\phi P_\mu P_\mu$.
Therefore, $\left\{ P_\phi, P_\mu  \right\}$ is not a code under matrix multiplication.
\end{xmpl}

The similarity between the following proposition and Lemma~\ref{lem:det-nonzero-twotwo}.$(i)$ 
shows 
further similarity between $\hom(\FreeM)$ and $\carre{\N}{2}$.

\begin{prpstn}
Let $\sigma \in \hom(\FreeM)$ be such that $\sigma$ is non-injective.
For every $\tau \in \hom(\FreeM)$, equality $\sigma \sigma \tau \sigma = \sigma\tau \sigma \sigma$ holds.
\end{prpstn}

\begin{proof}
Put 
$\alpha \defeq \sigma \sigma  \tau \sigma$
and
$\beta \defeq \sigma \tau \sigma  \sigma$.
Since $\sigma$ is non-injective, there exist $s \in \FreeM$ and $p$, $q \in \N$ such that $\sigma(\ze) = s^p$ and $\sigma(\on) = s^q$ (see Example~\ref{ex:zeon-d}).

First, $P_\sigma$ is singular because 
$$
P_\sigma
=
\twotwo
{p\nocc{\ze}{s}}{q\nocc{\ze}{s}}
{p\nocc{\on}{s}}{q\nocc{\on}{s}} 
=
\twoone{\nocc{\ze}{s}}{\nocc{\on}{s}}
\onetwo{p}{q}
\,.
$$
Therefore, 
we have 
$P_\sigma P_\sigma  P_\tau P_\sigma = P_\sigma P_\tau  P_\sigma P_\sigma$
by Lemma~\ref{lem:det-nonzero-twotwo}.$(i)$, 
and thus Claim~\ref{claim:mat-morph}.$(i)$ ensures that
$P_{\alpha} = P_{\beta}$.

Second, let $x \in \FreeM$.
For every $\rho \in \hom(\FreeM)$, we have 
$$
P_\rho \twoone{\nocc{\ze}{x}}{\nocc{\on}{x}} = \twoone{\nocc{\ze}{\rho(x)}}{\nocc{\on}{\rho(x)}} \,,
$$
and thus 
$$
\onetwo{1}{1} P_\rho \twoone{\nocc{\ze}{x}}{\nocc{\on}{x}} = \lgr{\rho(x)} \, .
$$
In particular, 
the latter equality holds for $\rho = \alpha$ and $\rho = \beta$, so $\lgr{\alpha (x)} = \lgr{\beta(x)}$.
Since $\sigma$ maps each element of $\FreeM$ to a power of $s$, 
we finally get
$\alpha (x) =  s^{\lgr{\alpha(x)} \lgr{s}^{-1}} = s^{\lgr{\beta(x)} \lgr{s}^{-1}} = \beta (x)$.
\end{proof}

\section{Three-by-three matrices} \label{sec:three-three}

The aim of this section is to prove that, for every integer $k \ge 13$, 
both  
\pbkfree{k}{\zeonzeon}
and 
\pbkfree{k}{\carre{\N}{3}} are undecidable.

We first  check that  ${\zeonzeon}$ is a well-behaved semigroup in the sense of Section~\ref{sec:nbGene}.

\begin{prpstn} \label{prop:monotone-prod}
Let $k_0$ be a positive integer.
If \pbkfree{k_0}{\zeonzeon} is decidable then for every $k \in \seg{1}{k_0}$, \pbkfree{k}{\zeonzeon} is also decidable.
\end{prpstn} 

\begin{proof}
 Let $u$, $v$, $w \in \FreeM$ be such that $\{ u, v, w \}$ is a $3$-element code. 
For instance, $\on $, $\ze \on$, and $\ze \ze \on$ are suitable choices for $u$, $v$, and $w$, respectively.  
Let $\sigma\colon \FreeM \to \FreeM$ be the morphism defined by:
 $\sigma(\ze) \defeq  u$ and $\sigma(\on) \defeq  v$.
For any $k$-element subset $X \subseteq \zeonzeon$, 
$$Y \defeq \left\{ (\sigma(x), \sigma(x')) : (x, x') \in X \right\} \cup \{ (w, w) \}$$
is a $(k + 1)$-element subset of $\zeonzeon$ that satisfies:
$X$ is a code \textiff{} $Y$ is a code.
Hence, there exists a many-one reduction from 
\pbkfree{k}{\zeonzeon} to \pbkfree{k + 1}{\zeonzeon}.
\end{proof}

Note that we do not know whether Proposition~\ref{prop:monotone-prod} still holds if $\zeonzeon$ is replaced with $\carre{\N}{3}$.

\subsection{Semi-Thue systems and Post correspondence problem}
\label{sec:semi-Thue-PCP}

In this section, we revisit Claus's reduction from the accessibility problem for semi-Thue systems to the Post correspondence problem \cite{Claus80}.
Semi-Thue systems are introduced in Definition~\ref{def:semi-Thue}.

\begin{dfntn}[Accessibility problem for semi-Thue systems \cite{Post47Thue}] \label{def:access}
Let \pbaccess{} denote the following problem: 
given a finite alphabet $\Sigma$, 
a subset $R \subseteq \Sigma^\star \times \Sigma^\star$, 
and two words $u$, $v \in \Sigma^\star$,
decide whether $v$ is accessible from $u$ under the semi-Thue system $(\Sigma, R)$.
For every integer $k \ge 1$, 
\pbkaccess{k} denotes the restriction of \pbaccess{} to those instances $(\Sigma, R, u, v)$ such that the cardinality of $R$ equals $k$.
\end{dfntn}

\begin{dfntn}[Post correspondence problem  \cite{Post46PCP}] \label{def:PCP}
Let \pbPCP{} denote the following problem: given 
a finite alphabet $\Sigma$ and 
two morphisms $\sigma$, $\tau\colon \Sigma^\star \to \FreeM$, 
decide whether there exists $w \in \Sigma^+$ such that $\sigma(w) = \tau(w)$. 
For every integer $k \ge 1$,
 \pbkPCP{k} denotes the restriction of \pbPCP{} to those instances $(\Sigma, \sigma, \tau)$ such that 
the cardinality of $\Sigma$ equals $k$.
\end{dfntn} 

\begin{rmrk}
Strictly speaking, 
\pbPCP{} is not a restriction of \pbGPCP. 
However, 
there is a simple, natural reduction from \pbPCP{} to \pbGPCP: 
for any instance  $(\Sigma, \sigma, \tau)$ of \pbPCP, 
$(\Sigma, \sigma, \tau)$ is a yes-instance of \pbPCP{} 
\textiff{} 
there exists $a \in \Sigma$ such that $(\Sigma, \sigma, \tau, \sigma(a), \mv, \tau(a), \mv)$ is a yes-instance of \pbGPCP.
An even more natural idea is to transform  $(\Sigma, \sigma, \tau)$  into $(\Sigma,  \sigma, \tau, \mv, \mv, \mv, \mv)$, but unfortunately, $(\Sigma,  \sigma, \tau, \mv, \mv, \mv, \mv)$ is always a yes-instance   of \pbGPCP{} because $\sigma(\mv) = \mv= \tau(\mv)$.
\end{rmrk}

Let $\Sigma$ and $\Gamma$ be two finite alphabets and let $\sigma$, $\tau\colon \Sigma^\star \to \Gamma^\star$ be two morphisms.
Stricly speaking, $(\Sigma, \sigma, \tau)$ is not an instance of \pbPCP,
unless $\Gamma =  \zeon$.
However, we abuse language by identifying $(\Sigma, \sigma, \tau)$ with 
$(\Sigma, \gamma \sigma, \gamma \tau)$, 
where  $\gamma\colon \Gamma^\star \to \FreeM$ is any injective morphism (see Claim~\ref{claim:inject-zeon}). 

Post proved the undecidabilities of \pbPCP{} and \pbaccess{} in 1946 and 1947, respectively \cite{Post46PCP,Post47Thue}.
Since then, his results have been tremendously refined:

\begin{thrm}[Matiyasevich and S{\'e}nizergues 1996 \cite{MatiyasevichS05}] \label{th:access-trois}
\pbkaccess{3} is undecidable.
\end{thrm}

\begin{thrm}[Claus 1980 \cite{Claus80}] \label{th:Thue-Claus-orig}
Let $k$ be a positive integer.
If \pbkPCP{k + 4} is decidable then \pbkaccess{k} is decidable. 
\end{thrm}

It follows from Theorems~\ref{th:access-trois} and~\ref{th:Thue-Claus-orig}
that \pbkPCP{7} is undecidable \cite{MatiyasevichS05}. 
To complete the picture, 
let us mention that 
\pbkPCP{2} is decidable \cite{HalavaHH02}, and that the decidabilities of 
\pbkaccess{1}, 
\pbkaccess{2}, 
\pbkPCP{3},  
\pbkPCP{4}, 
\pbkPCP{5}, and  
\pbkPCP{6}  remain open. 

To prove Theorem~\ref{th:Thue-Claus-orig}, Claus presents a many-one reduction from \pbkaccess{k}  to \pbkPCP{k + 4} \cite{Claus80} (similar proofs can be found in \cite{Claus07,Nicolas08,HarjuKHandbook}).
In 2007,
 Halava, Harju, and Hirvensalo remarked that Claus's construction is freeness-friendly.
In fact, as we shall see,
it turns out that for any instance $(\Sigma, \sigma, \tau)$ of \pbPCP{} computed by the reduction, 
$(\Sigma, \sigma, \tau)$ is a yes-instance of \pbPCP{} \textiff{}
 $\left\{ (\sigma(a), a) : a \in \Sigma \right\} \cup \left\{ (\tau(a), a) : a \in \Sigma \right\}$ is not a code under  componentwise concatenation.

\begin{dfntn}[Claus instance of \pbPCP] \label{def:ClausInstance}
Define a \emph{Claus instance} of \pbPCP{}  as a triple of the form $(\Sigma, \sigma, \tau)$, where $\Sigma$ is a finite alphabet and $\sigma$, $\tau\colon \Sigma^\star \to \zobed^\star$ are morphisms meeting the requirements listed below:
\begin{itemize}
\item 
$\ttb \in \Sigma$, 
$\tte \in \Sigma$,  
\item 
$\sigma(a) \in  {\{ \ttd \ze,  \ttd \on \}}^+$ for every $a \in \Sigma \setminus \{ \ttb,  \tte \}$,
\item 
$\tau(a) \in  {\{ \ze \ttd, \on\ttd \}}^+$ for  every $a \in \Sigma \setminus \{ \ttb, \tte \}$,
\item
$\sigma(\ttb)  \in  \ttb {\{ \ttd \ze,  \ttd \on \}}^\star$,
\item
$\sigma(\tte) \in     {\{ \ttd \ze,  \ttd \on \}}^\star \ttd \tte$, 
\item 
$\tau(\ttb) \in \ttb \ttd {\{ \ze \ttd, \on\ttd \}}^\star$,  and 
\item 
$\tau(\tte) \in  {\{ \ze \ttd, \on\ttd \}}^\star \tte$.
\end{itemize}
\end{dfntn}

Strictly speaking, 
Claus's original reduction \cite{Claus80} does not  output Claus instances in the sense of Definition~\ref{def:ClausInstance}, 
but it can be easily adapted.
Other similar constructions \cite{Claus07,HarjuKHandbook} are also  adaptable. 

\begin{thrm}[Claus's theorem revisited] 
 \label{thm:Thue-Claus}
Let $k$ be a positive integer.
If \pbkPCP{k + 4} is decidable on Claus instances then \pbkaccess{k} is decidable.
\end{thrm}

A full proof Theorem~\ref{thm:Thue-Claus}  can be found in an unpublished paper by the second author \cite{Nicolas08}. 
Note that Theorem~\ref{thm:Thue-Claus} is a corollary of the following two facts:
\begin{enumerate}
\item if \pbkGPCP{k + 2} is decidable then \pbkaccess{k} is decidable and 
\item if \pbkPCP{k + 2} is decidable on Claus instances then \pbkGPCP{k} is decidable \cite{HarjuKHandbook}. 
\end{enumerate} 
Combining Theorems~\ref{th:access-trois} and~\ref{thm:Thue-Claus}  yields:

\begin{crllr} \label{cor:PCP-Claus}
\pbkPCP{7} is undecidable on Claus instances.
\end{crllr}

\subsection{Mixed modification of the Post correspondence problem} \label{sec:MMPCP}

The following problem is a useful hybrid between  \pbfree{\zeonzeon} and \pbPCP.

\begin{dfntn}[Mixed Modification of the \pbPCP{} \cite{CassaigneHK99}]
Let \pbMMPCP{} denote the following problem:
given an instance $(\Sigma, \sigma, \tau)$ of \pbPCP,
decide whether  there exist an integer $n \ge 1$, $n$ symbols $a_1$, $a_2$, \ldots, $a_n \in \Sigma$ and $2n$ morphisms 
$\sigma_1$, $\sigma_2$, \ldots,  $\sigma_n$, $\tau_1$, $\tau_2$, \ldots, $\tau_n \in \{ \sigma, \tau \}$ 
such that 
\begin{equation} \label{eq:sigma-eq-tau}
\sigma_1(a_1)\sigma_2(a_2) \dotsm \sigma_n(a_n) = \tau_1(a_1)\tau_2(a_2) \dotsm \tau_n(a_n) 
\end{equation}
and 
\begin{equation} \label{eq:sigma-neq-tau}
 (\sigma_1, \sigma_2, \dotsc,  \sigma_n) \ne (\tau_1, \tau_2, \dotsc, \tau_n) \, . 
\end{equation}
For every integer $k \ge 1$,
 \pbkMMPCP{k} denotes the restriction of \pbMMPCP{} to those instances $(\Sigma,  \sigma, \tau)$ such that the cardinality of  $\Sigma$ equals $k$.
\end{dfntn}

The fundamental property of \pbMMPCP{} can be stated as follows:

\begin{clm} \label{claim:MMPCP-freeprod}
Let  $(\Sigma, \sigma, \tau)$  be an instance of \pbMMPCP{} such that $\sigma(a) \ne \tau(a)$ for every $a \in \Sigma$.
$(\Sigma, \sigma, \tau)$ is a yes-instance of \pbMMPCP{} \textiff{}  
$\left\{ (\sigma(a), a) : a \in \Sigma \right\} \cup \left\{ (\tau(a), a) : a \in \Sigma \right\}$ is not a code under componentwise concatenation.
\end{clm}

It is clear that every yes-instance of \pbPCP{} is also a yes-instance of \pbMMPCP{}. 
The following proposition ensures that the converse is true for  Claus instances.

\begin{prpstn} \label{lem:claus}
For every Claus instance $(\Sigma,  \sigma, \tau)$ of  \pbPCP,
$(\Sigma,  \sigma, \tau)$ is a yes-instance of \pbMMPCP{} \textiff{}  there exists $w \in {(\Sigma \setminus \{ \ttb, \tte \})}^\star$ such that $\sigma(\ttb w \tte) =  \tau(\ttb w \tte)$.
\end{prpstn}

\begin{proof}
The ``if part'' is trivial. Let us now prove the ``only if part''.

For each integer $n \ge 1$, 
define $\calc_n$ as the set of those $n$-tuples $\left( a_i, \sigma_i, \tau_i \right)_{i \in \seg{1}{n}}$ over $\Sigma \times \{ \sigma, \tau \} \times \{ \sigma, \tau \}$  that satisfy Equation~\eqref{eq:sigma-eq-tau} 
and define $\bcalc_n$ as the set of those $\left( a_i, \sigma_i, \tau_i \right)_{i \in \seg{1}{n}} \in \calc_n$ that satisfy Condition~\eqref{eq:sigma-neq-tau}.
There exists an integer $n \ge 1$ such that  $\bcalc_n \ne \emptyset$
\textiff{}
$(\Sigma, \{ \ze, \on,  \ttb, \tte, \ttd \}, \sigma, \tau)$ is a yes-instance of \pbMMPCP.

\begin{clm} \label{claim:tau1-sigma1}
For any $\left( a_i, \sigma_i, \tau_i \right)_{i \in \seg{1}{n}} \in \bcalc_n$,  $\sigma_1 = \tau_1$  implies $\left( a_i, \sigma_i, \tau_i \right)_{i \in \seg{2}{n}} \in \bcalc_{n - 1}$. 
\end{clm} 

\begin{clm}\label{claim:taun-sigman}
For any $\left( a_i, \sigma_i, \tau_i \right)_{i \in \seg{1}{n}} \in \bcalc_n$,  $\sigma_n = \tau_n$  implies $\left( a_i, \sigma_i, \tau_i \right)_{i \in \seg{1}{n - 1}} \in \bcalc_{n - 1}$. 
\end{clm}

\begin{lmm} \label{lem:b-e-milieu}
Let $\left( a_i, \sigma_i, \tau_i \right)_{i \in \seg{1}{n}} \in \bcalc_n$ and let $k \in \seg{1}{n - 1}$.
If $a_k = \tte$ or $a_{k + 1} = \ttb$ then 
$\left( a_i, \sigma_i, \tau_i \right)_{i \in \seg{1}{k}}$ belongs to $\bcalc_k$ or 
$\left( a_i, \sigma_i, \tau_i \right)_{i \in \seg{k + 1}{n}}$ belongs to $\bcalc_{n - k}$.
\end{lmm} 

\begin{proof} 
We only prove the statement in the case where $a_k = \tte$
 because the case where $a_{k + 1} = \ttb$ is handled in the same way.

Put 
$s \defeq \sigma_1(a_1) \sigma_2(a_2) \dotsm \sigma_k(a_k)$ and 
$t \defeq \tau_1(a_1) \tau_2(a_2) \dotsm \tau_k(a_k)$. 
We have 
$\nocc{\tte}{\sigma(\tte)} = \nocc{\tte}{\tau(\tte)} = 1$
and
$\nocc{\tte}{\sigma(a)} = \nocc{\tte}{\tau(a)} = 0$ for every $a \in \Sigma \setminus \{ \tte \}$.
From that we deduce:
\begin{equation} \label{eq:s-t-e}
\nocc{\tte}{s} 
= \nocc{\tte}{a_1 a_2 \dotsm a_k}  
= \nocc{\tte}{t} \, .
\end{equation} 
Assume that $a_k = \tte$.
Now,  both   $s$ and $t$ end with $\tte$.
Hence, if  $t$ was a proper prefix of  $s$ then we would have 
$\nocc{\tte}{t} < \nocc{\tte}{s}$ in contradiction with Equation~\eqref{eq:s-t-e}.
In the same way $s$ cannot be a proper prefix of  $t$.  
Therefore, $s$ equals $t$.
It follows 
that
$\left( a_i, \sigma_i, \tau_i \right)_{i \in \seg{1}{k}} \in \calc_k$
and 
$\left( a_i, \sigma_i, \tau_i \right)_{i \in \seg{k + 1}{n}} \in \calc_{n - k}$,
and at least one of them satisfies Condition~\eqref{eq:sigma-neq-tau}. 
\end{proof} 

Let $n$ be the smallest positive integer such that   $\bcalc_n \ne \emptyset$;
let $\left( a_i, \sigma_i, \tau_i \right)_{i \in \seg{1}{n}}$ be an element of $\bcalc_n$.
 Claim~\ref{claim:tau1-sigma1} ensures $\sigma_1 \ne \tau_1$,
and since  $\sigma_1(a_1)$ and $\tau_1(a_1)$ start with the same letter, 
we have $a_1 = \ttb$.
In the same way, Claim~\ref{claim:taun-sigman} ensures $\sigma_n \ne \tau_n$, and since  $\sigma_n(a_n)$ and  $\tau_n(a_n)$ end with the same letter, we have
$a_n = \tte$.
Furthermore, 
Lemma~\ref{lem:b-e-milieu} ensures
that $w \defeq a_2 a_3 \dotsm a_{n - 1}$  belongs to ${(\Sigma \setminus \{ \ttb, \tte \})}^\star$.

Without loss of generality, we may assume $\sigma_1 = \sigma$ and $\tau_1 = \tau$. 
To complete the proof of the proposition, 
it suffices to show that, 
for every $i \in \seg{2}{n}$,  $\sigma_i = \sigma$ and $\tau_i = \tau$.
We proceed by induction.
Let $i$, $j \in \seg{1}{n}$ be such that 
$\sigma = \sigma_1 = \sigma_2 =  \cdots = \sigma_i$ and 
$\tau = \tau_1 = \tau_2 =  \cdots = \tau_j$.
\begin{enumerate}[$(i)$.]
\item If $\sigma(a_1 a_2 \dotsm a_i) = \tau(a_1 a_2 \dotsm a_j)$ then 
$a_1 a_2 \dotsm a_i = a_1 a_2 \dotsm a_j = \ttb w \tte$.

Indeed,
 if $\sigma(a_i)$ and $\tau(a_j)$ end with the same letter,
then $a_i = a_j = \tte$ and $i = j = n$ follows.

\item If  $\sigma(a_1 a_2 \dotsm a_i)$ is a proper  prefix of $\tau(a_1 a_2 \dotsm a_j)$ then $\sigma_{i + 1} = \sigma$.

Indeed, assume that there exists a non-empty word $s$ such that $\sigma(a_1 a_2 \dotsm a_i) s = \tau(a_1 a_2 \dotsm a_j)$.
On the one hand, $s$ starts with the same letter as $\sigma_{i + 1}(a_{i + 1})$.
On the other hand, $s$ belongs to ${\{ \ttd \ze,  \ttd \on \}}^\star \ttd$
since  $\sigma(a_1 a_2 \dotsm a_i) \in \ttb {\{ \ttd \ze,  \ttd \on \}}^\star$ while 
 $\tau(a_1 a_2 \dotsm a_j) \in \ttb \ttd {\{ \ze \ttd, \on\ttd \}}^\star$.
 Hence, $\sigma_{i + 1}(a_{i + 1})$ starts with $\ttd$, and $\sigma_{i + 1} = \sigma$ follows.

\item If $\tau(a_1 a_2 \dotsm a_j)$ is a proper prefix of   $\sigma(a_1 a_2 \dotsm a_i)$ then $\tau_{j + 1} = \tau$. 

Point~$(iii)$ is proven in the same way as point~$(ii)$.
\end{enumerate}
\end{proof}

\begin{thrm}[Halava, Harju, and Hirvensalo 2007 \cite{Claus07}] \label{thm-MMPCP-Claus}
\pbkMMPCP{7} is undecidable on Claus instances.
\end{thrm} 

\begin{proof}
Proposition~\ref{lem:claus} ensures that  \pbPCP{} and  \pbMMPCP{} are equivalent on Claus instances.
Therefore, \pbkMMPCP{7} is undecidable on Claus instances by Corollary~\ref{cor:PCP-Claus}.
\end{proof} 

Note that the decidability of  \pbkMMPCP{k} remains open for each $k \in \seg{2}{6}$.

\subsection{Proofs of the main results}

We first prove that  \pbkfree{k}{\zeonzeon} is undecidable for every integer $k \ge 13$.
The idea is to construct  a  many-one reduction from  \pbkMMPCP{7} on Claus instances to \pbkfree{13}{\zeonzeon}.

\begin{lmm} \label{lem:code-st12}
Let $S$ be a semigroup, 
let $X$ be a subset of $S$, 
let $s_1$, $t_1$, $s_2$, $t_2 \in X$, and 
let $Y \defeq (X \setminus \{ s_1, t_1, s_2, t_2 \}) \cup \{ t_2 s_1, s_2 t_1, t_2 t_1 \}$.
If $X$ is a code then $Y$ is also a code.
\end{lmm}

\begin{proof}
If $X$ is an alphabet and if $S = X^+$ then $Y$ is a prefix code over $X$.
The general case  follows.
\end{proof}

The converse of Lemma~\ref{lem:code-st12} is false in general.
For instance, consider the case where $S \defeq {\{ \ze, \on, \tw \}}^+$, 
$s_1 \defeq \ze \on$,
$t_1 \defeq  \tw$, 
$s_2 \defeq \ze$ 
$t_2 \defeq \on \tw$,
and $X \defeq  \{ s_1, t_1, s_2, t_2 \}$.
Then, $Y =  \{ t_2 s_1, s_2 t_1, t_2 t_1 \} = \{ \on \tw  \ze \on, \ze  \tw, \on \tw  \tw \}$ is a prefix code,
but $X$ is not a code because $s_1t_1 =  \ze \on \tw = s_2 t_2$.

\begin{thrm} \label{th:zeonzeon}
Let $k$ be a positive integer.
If \pbkfree{2k - 1}{\zeonzeon} is decidable then both \pbkPCP{k} and \pbkMMPCP{k} are decidable on Claus instances.
\end{thrm}

\begin{proof}
Let $(\Sigma, \sigma, \tau)$ be a Claus  instance of \pbkPCP{k}. 
For each $w \in \Sigma^\star$, put 
$$s_w  \defeq (\sigma(w), w)$$
 {and} 
$$t_w  \defeq (\tau(w), w) \, .$$
Let $X$ and $Y$ be the two subsets of $\zobed^\star \times \Sigma^\star$ defined by:
$$
X  \defeq \left\{ s_a : a \in \Sigma \right\} \cup \left\{ t_a  : a \in \Sigma \right\} 
$$
and 
$$
Y  \defeq 
\left(  X \setminus \left\{ s_\ttb, t_\ttb, s_\tte, t_\tte  \right\} \right) 
\cup 
\left\{ t_\tte s_\ttb, s_\tte t_\ttb,  t_\tte t_\ttb \right\} \, . 
$$
Let $\phi\colon \Sigma^\star \to \FreeM$ and $\psi\colon  \zobed^\star \to \FreeM$ be two injective morphisms (see Claim~\ref{claim:inject-zeon}),
 and let 
$$Z \defeq \left\{ (\psi(y), \phi(y')) : (y, y') \in Y  \right\} \, . 
$$

It is clear that $X$, $Y$, and $Z$ are computable from  $(\Sigma,  \sigma, \tau)$.
Moreover, 
the cardinality of $X$ equals $2k$,  
the cardinality of $Y$ equals $2k - 1$, and 
$Z$ is an instance of  \pbkfree{2k - 1}{\zeonzeon}.
It remains to prove that the following five assertions are equivalent:
\begin{enumerate}[$(i)$.]
\item \label{item-PCP-yes}
$(\Sigma, \sigma, \tau)$ is a yes-instance of  \pbkPCP{k}.
\item \label{item-MMPCP-yes}
$(\Sigma, \sigma, \tau)$ is a yes-instance of  \pbkMMPCP{k}.
\item \label{item-X-not-code}
$X$ is not a code. 
\item \label{item-Y-not-code}
$Y$ is not a code. 
\item \label{item-Z-not-code}
$Z$ is not a code.
\end{enumerate}

By Proposition~\ref{lem:claus}, $(\ref{item-PCP-yes})$ and $(\ref{item-MMPCP-yes})$ are equivalent.
By Claim~\ref{claim:MMPCP-freeprod}, $(\ref{item-MMPCP-yes})$ and $(\ref{item-X-not-code})$ are equivalent.
By Lemma~\ref{lem:code-st12}, $(iv)$ implies $(iii)$.
Since $Z$ is the image of $Y$ under an injective morphism, $(\ref{item-Y-not-code})$ and $(\ref{item-Z-not-code})$ are equivalent.

Assume that $(\Sigma, \sigma, \tau)$ is a yes-instance of   \pbkMMPCP{k}.
By Proposition~\ref{lem:claus}, there exists  $w \in {(\Sigma \setminus \{ \ttb, \tte \})}^\star$ such that $\sigma(\ttb w \tte) =  \tau(\ttb w \tte)$.
The word $w$ satisfies $s_\ttb s_w s_\tte = s_{\ttb w \tte} = t_{\ttb w \tte} = t_\ttb t_w t_\tte$,
 and  thus we have
\begin{equation} \label{eq:Y-not-a-code}
(t_ \tte s_\ttb) s_w (s_\tte t_\ttb)  =  ( t_\tte t_\ttb) t_w (t_\tte t_\ttb) \, .
\end{equation}
Since  $t_ \tte s_\ttb \in Y$, $t_\tte t_\ttb \in Y$, $ s_w (s_\tte t_\ttb) \in Y^+$, $t_w (t_\tte t_\ttb) \in Y^+$, and $t_ \tte s_\ttb \ne t_\tte t_\ttb$,
it follows from Equation~\eqref{eq:Y-not-a-code} that $Y$ is not a code. 
Therefore,  $(ii)$ implies $(iv)$.
\end{proof} 


\begin{crllr} \label{cor:zeonzeon}
For every integer $k \ge 13$, 
\pbkfree{k}{\zeonzeon} is undecidable.
\end{crllr}

\begin{proof}
Combining Corollary~\ref{cor:PCP-Claus} and Theorem~\ref{th:zeonzeon},
 we obtain that
\pbkfree{13}{\zeonzeon} is undecidable.
Hence, the corollary holds by Proposition~\ref{prop:monotone-prod}.
\end{proof}

By way of digression, 
let us briefly summarize the current knowledge  about the decidability of 
\pbkfree{k}{\FreeM^{\times d}} for $k$, $d \in \N \setminus  \{ 0 \}$.
On the one hand, \pbkfree{k}{\FreeM} is decidable for every integer $k \ge 1$ \cite{BerstelPR09},
and so is \pbkfree{2}{\FreeM^{\times d}} for every integer $d \ge 1$ (Example~\ref{ex:zeon-d}).
On the other hand, if 
\pbkfree{k}{ \FreeM^{\times (d + 1)}} 
is decidable then 
\pbkfree{k}{\FreeM^{\times d}} 
is also decidable because there exist injective morphisms from  $\FreeM^{\times d}$ to $\FreeM^{\times (d + 1)}$:
for instance, the function mapping each $(u_1, u_2, \dotsc, u_d) \in \FreeM^{\times d}$ to $(u_1, u_2, \dotsc, u_d, \mv)$.
Hence,  
it follows from Corollary~\ref{cor:zeonzeon} that 
\pbkfree{k}{\FreeM^{\times d}} is undecidable for 
every $(k, d) \in (\N \setminus \seg{0}{12}) \times (\N \setminus \{ 0, 1 \})$.

\begin{open} 
For each $(k, d) \in \seg{3}{12} \times (\N \setminus \{ 0, 1 \})$,
the decidability of \pbkfree{k}{\FreeM^{\times d}} is open.
\end{open} 

Let us now return to our main plot.
It remains to prove that \pbkfree{k}{\carre{\N}{3}} is undecidable for every integer $k \ge 13$.

\begin{lmm}[\cite{Paterson70,CassaigneHK99,Claus07}] \label{lem:inject-three-three}
There exists an injective morphism from $\zeonzeon$ to $\carre{\N}{3}$.
\end{lmm}

\begin{proof}
Let  $\beta\colon \FreeM \to \N$ be defined by:
$\beta(\ze) = 0$, 
$\beta(\on) = 1$,
and 
$\beta(uv) =  \beta(u) + 2^{\lgr{u}} \beta(v)$ for every $u$, $v \in \FreeM$.
The word $a_n \dotsm a_2 a_1$  is a  binary expansion of the natural number $\beta(a_1 a_2 \dotsm a_n)$ for any integer $n \ge 1$ and any $a_1$, $a_2$, \ldots, $a_n \in \zeon$.
Let $\Phi\colon \zeonzeon \to \carre{\N}{3}$ be defined by:
$$
\Phi(u, v)  \defeq 
\begin{bmatrix}
2^{\lgr{u}} & 0 & \beta(u) \\
0 & 2^{\lgr{v}} & \beta(v) \\
0 & 0 & 1 
\end{bmatrix}
$$
for every  $u$, $v   \in \FreeM$.
It is easy to check that $\Phi$ is a morphism:
 $\Phi(u u', v v') = \Phi(u, v) \Phi(u', v')$ for all $u$, $u'$, $v$, $v' \in \FreeM$.
Note that  $\beta$ is not injective since $\beta( u) = \beta(u\ze)$ for every $u \in \FreeM$.
However,  the function mapping each $u \in \FreeM$ to $(\lgr{u}, \beta(u))$ is injective, so  $\Phi$ is injective.
\end{proof} 

Lemma~\ref{lem:inject-three-three} can be easily generalized to higher dimensions:  
for every integer $d \ge 1$, 
there exists an injective morphism from $\FreeM^{\times d}$ to $\carre{\N}{(d + 1)}$.
However, there is no injective injective morphism from $\zeonzeon$ to $\carre{\C}{2}$ \cite{CassaigneHK99}. 

\begin{thrm} \label{th:N3-zeonzeon}
Let $k$ be a positive integer.
If \pbkfree{k}{\carre{\N}{3}}  is decidable then \pbkfree{k}{\zeonzeon} is decidable.
\end{thrm}

\begin{proof}
Any injective morphism from $\zeonzeon$ to $\carre{\N}{3}$ induces a one-one reduction from \pbkfree{k}{\zeonzeon} to  \pbkfree{k}{\carre{\N}{3}}.
Hence, the desired result follows from Lemma~\ref{lem:inject-three-three}.
\end{proof}

From  Corollary~\ref{cor:zeonzeon} and Theorem~\ref{th:N3-zeonzeon} we deduce:

\begin{crllr} \label{cor:three-three-undec}
For every integer $k \ge 13$, \pbkfree{k}{\carre{\N}{3}} is undecidable.
\end{crllr}

\section{Matrices of higher dimension}
\label{sec:deux-matrices}

The main aim of this section is to prove that \pbkfree{2}{\carre{\N}{d}} is undecidable for some integer $d \ge 1$.
Although the result is not new \cite{Mazoit98}, 
it has never been published before.

\begin{thrm} \label{th:reduc-dim}
Let $D$ be a semiring with a \ruset{} and let $k$ and $d$ be positive integers.
If  
\pbkfree{k + 1}{\carre{D}{d}} is decidable 
then 
\pbkfree{kd + 1}{D} is decidable.
\end{thrm} 

\begin{proof}
We present a many-one reduction from \pbkfree{kd + 1}{D} to \pbkfree{k + 1}{\carre{D}{d}}.
The construction is the same as in \cite{Mazoit98}.
For each $n \in \N$, let $I_n$ denote the $n$-by-$n$ identity matrix over $D$.

Let $X$ be a $(kd + 1)$-element subset of $D$.
Write $X$ in the form
$$
X = \left\{ a  \right\}
\cup \left\{ b_{i,j} : (i, j) \in \seg{1}{d} \times \seg{1}{k}\right\} \, .
$$
For each $j \in \seg{1}{k}$, 
let $B_j$ denote the $d$-by-$d$ matrix over $D$ given by: 
 the rightmost column of $B_j$ equals the transpose of 
 $$
 \begin{bmatrix}
   b_{d,j} &  \cdots &  b_{3, j}  &  b_{2, j} & b_{1, j} 
 \end{bmatrix}
 $$ 
 and 
all entries of $B_j$ that are not located in its rightmost column equal zero.
Put 
$$
A \defeq \twotwo{O}{a}{I_{d-1}}{O} 
$$
and 
$$
\calx \defeq \left\{ A, B_1, B_2, \dotsc, B_{k} \right\} \, .
$$ 
For instance in the case where $d = 4$ and $k = 3$, we have:
$$
\calx = 
\left\{ 
 \begin{bmatrix}
 0 & 0 & 0 & a \\
 1 & 0 & 0 & 0 \\
 0 & 1 & 0 & 0 \\
 0 & 0 & 1 & 0 
 \end{bmatrix},
 \begin{bmatrix}
 0 & 0 & 0 & y_{4,1} \\
 0 & 0 & 0 & y_{3,1} \\
 0 & 0 & 0 & y_{2,1} \\
 0 & 0 & 0 & y_{1,1}  
 \end{bmatrix},
 \begin{bmatrix}
 0 & 0 & 0 & y_{4,2} \\
 0 & 0 & 0 & y_{3,2} \\
 0 & 0 & 0 & y_{2,2} \\
 0 & 0 & 0 & y_{1,2}  
 \end{bmatrix},
\begin{bmatrix}
 0 & 0 & 0 & y_{4,3} \\
 0 & 0 & 0 & y_{3,3} \\
 0 & 0 & 0 & y_{2,3} \\
 0 & 0 & 0 & y_{1,3}  
 \end{bmatrix}
 \right\}  \, .
$$
Clearly, $\calx$  is a $(k + 1)$-element subset of $\carre{D}{d}$ and $\calx$ is computable from $X$.
To complete the proof of the theorem, 
it remains to check that
$X$ is a code under the multiplicative operation of $D$ 
\textiff{}  
$\calx$ is a code under the matrix multiplication induced by the operations of $D$.

Let $\calc$ denote the following instance of the gadget introduced in Definition~\ref{def:gadget-nbgene}:
$$
\calc \defeq
C_d\left(A, \left\{ B_1, B_2, \dotsc, B_{k} \right\} \right)
=
\left\{ A^d \right\} \cup 
\left\{ A^{i - 1} B_j : (i, j) \in \seg{1}{d} \times \seg{1}{k} \right\} \, .
$$
Let $\phi\colon \carre{D}{d} \to D$ be defined by: 
for every $M \in \carre{D}{d}$, 
$\phi(M)$ equals the ${(d, d)}^\text{th}$ entry of $M$.
Recall from Section~\ref{sec:two-up} that $\Tri(d, D)$ denotes the semiring of $d$-by-$d$ upper-triangular matrices over $D$.

\begin{lmm} \label{lem:phi-induc}  \leavevmode 
 \begin{enumerate}[$(i)$.]
  \item $\phi$ induces a morphism from $\Tri(d, D)$ to $D$.
  \item $\calc$ is a subset $\Tri(d, D)$.
  \item $\phi$ induces a bijection from $\calc$  onto $X$.
 \end{enumerate}
\end{lmm}

\begin{proof}
Part~$(i)$ is clear: it simply means that $\phi(T T') = \phi(T) \phi(T')$ for all $T$, $T' \in \Tri(d, D)$.
Let us simultaneously prove parts~$(ii)$ and~$(iii)$.
It is easy to see that 
$$
A^i =
\twotwo{O}{a I_i}{I_{d - i}}{O}
$$
for every  $i \in \seg{0}{d}$.
In particular, it holds that $A^d = a I_d$.
It follows that  $A^d \in \Tri(d, D)$ and $\phi(A^d) = a$.
Now, let $i \in \seg{1}{d}$ and $j \in \seg{1}{k}$.
All non-zero entries of $A^{i - 1}  B_j$ are located in its rightmost column, 
so $A^{i - 1} B_j \in \Tri(d, D)$;
it is easy to see that $\phi(A^{i - 1} B_j) = b_{i, j}$.
\end{proof}

%

%

By Lemma~\ref{lem:phi-induc}.$(iii)$, 
the cardinality of $\calc$ equals $kd + 1$,
and thus, 
by Lemma~\ref{lem:gadget-code}, 
$\calx$ is a code \textiff{} $\calc$ is a code.
 Moreover, 
combining Claim~\ref{clm:pre-image-code} and Lemma~\ref{lem:phi-induc}, 
we get that if $X$ is a code  then $\calc$ is a code.
Let us show that the converse is also true.
Equality $B_1 M =  B_1 \phi(M)$ holds for every $M \in \Tri(d, D)$.
It follows that for all  $M$, $M' \in \Tri(d, D)$, 
$\phi(M) = \phi(M')$ implies $B_1 M = B_1 M'$.
Assume that $\calc$ is a code.
Then, $B_1$ is cancellative in $\calc^+$.
Besides, the latter set is a subset of $\Tri(d, D)$ by Lemma~\ref{lem:phi-induc}.$(ii)$.
Hence, $\phi$ is injective on $\calc^+$.
It then follows from Claim~\ref{clm:pre-image-code} and Lemma~\ref{lem:phi-induc} that $X$ is a code.

The proof of the theorem is now complete because we have just shown that the following three assertions are equivalent:
$X$ is a code, 
$\calc$ is a code, 
and $\calx$ is a code.
\end{proof}

\begin{crllr} \label{cor:dim-freeness}
For every $h \in \N$, 
\pbkfree{7 + h}{\carre{\N}{6}}, 
\pbkfree{5 + h}{\carre{\N}{9}},
\pbkfree{4 + h}{\carre{\N}{12}},
\pbkfree{3 + h}{\carre{\N}{18}}, and 
\pbkfree{2 + h}{\carre{\N}{36}} 
are undecidable.
\end{crllr}

\begin{proof}
Let $k$ and $d$ be positive integers.
Apply Theorem~\ref{th:reduc-dim} with $D \defeq \carre{\N}{3}$ and identify 
$\carre{\left( \carre{\N}{3} \right)}{d}$ with $\carre{\N}{3d}$:
if  
\pbkfree{kd+1}{\carre{\N}{3}} is undecidable
then 
\pbkfree{k+1}{\carre{\N}{3d}} is undecidable.
Hence, Corollary~\ref{cor:dim-freeness}  
follows from Corollary~\ref{cor:three-three-undec}.
\end{proof} 

In particular, \pbkfree{2}{\carre{\N}{36}} is undecidable.

\begin{lmm} \label{lem:inj-d-dplusun}
For any semiring $D$ with a \ruset{} and any integer $d \ge 1$,
there exists a computable, injective morphism from $\carre{D}{d}$ to  $\carre{D}{(d + 1)}$.
\end{lmm}

\begin{proof}
Map  each $M \in \carre{D}{d}$ to $\twotwo{M}{O}{O}{1}$. 
\end{proof}

Let $k$ and $d$ be positive integers.
If  \pbkfree{k}{\carre{\N}{d}} is undecidable then it  follows from Lemma~\ref{lem:inj-d-dplusun} that
 \pbkfree{k}{\carre{\N}{e}} is undecidable for every integer $e \ge d$.
Table~\ref{tab:free} summarizes our results on  the decidability of \pbkfree{k}{\carre{\N}{d}}  as $(k, d)$ runs over $(\N \setminus \{ 0 \}) \times (\N \setminus \{ 0, 1 \})$.
The table is to be understood as follows: 
if the symbol that occurs  at the intersection of  row $d$ and column $k$  is a ``D'' then  \pbkfree{k}{\carre{\N}{d}} is  decidable,
if it is a ``U'' then the problem  is undecidable, 
and if it is a ``$\qmark$'' then the decidability of the problem is still open.
\begin{table}
 \scalebox{0.968}{
\begin{tabular}{|cc| ccc ccc ccc ccc ccc c}
\hline
& & $k$ \\
& & $1$ & $2$ & $3$ & $4$ & $5$ & $6$ & $7$ & $8$ & $9$ & $10$ & $11$ & $12$ & $13$ & $14$  & $15$  &  \\
\hline 
$d$ & 2  & {D} & $\qmark$ & $\qmark$ & $\qmark$ & $\qmark$ & $\qmark$ & $\qmark$ & $\qmark$ & $\qmark$ & $\qmark$ & $\qmark$ & $\qmark$ & $\qmark$ & $\qmark$ & $\qmark$ &$\cdots$ \\
& 3  & {D} & $\qmark$ & $\qmark$ & $\qmark$ & $\qmark$ & $\qmark$ & $\qmark$ & $\qmark$ & $\qmark$ & $\qmark$ & $\qmark$ & $\qmark$ & {U} & {U} & {U} &$\cdots$  \\
& 4  & {D} & $\qmark$ & $\qmark$ & $\qmark$ & $\qmark$ & $\qmark$ & $\qmark$ & $\qmark$ & $\qmark$ & $\qmark$ & $\qmark$ & $\qmark$ & {U} & {U} & {U} &$\cdots$ \\
& 5 & {D} & $\qmark$ & $\qmark$ & $\qmark$ & $\qmark$ & $\qmark$ & $\qmark$ & $\qmark$ & $\qmark$ & $\qmark$ & $\qmark$ & $\qmark$ & {U} & {U} & {U} &$\cdots$ \\
& 6  & {D} & $\qmark$ & $\qmark$ & $\qmark$ & $\qmark$ & $\qmark$ & {U} & {U} & {U} & {U} & {U} & {U} & {U} & {U} & {U} &$\cdots$ \\
& 7  & {D} & $\qmark$ & $\qmark$ & $\qmark$ & $\qmark$ & $\qmark$ & {U} & {U} & {U} & {U} & {U} & {U} & {U} & {U} & {U} &$\cdots$ \\
& 8  & {D} & $\qmark$ & $\qmark$ & $\qmark$ & $\qmark$ & $\qmark$ & {U} & {U} & {U} & {U} & {U} & {U} & {U} & {U} & {U} &$\cdots$ \\
& 9  & {D} & $\qmark$ & $\qmark$ & $\qmark$ & {U} & {U} & {U} & {U} & {U} & {U} & {U} & {U} & {U} & {U} & {U} &$\cdots$ \\
& 10 & {D} & $\qmark$ & $\qmark$ & $\qmark$ & {U} & {U} & {U} & {U} & {U} & {U} & {U} & {U} & {U} & {U} & {U} &$\cdots$ \\
& 11 & {D} & $\qmark$ & $\qmark$ & $\qmark$ & {U} & {U} & {U} & {U} & {U} & {U} & {U} & {U} & {U} & {U} & {U} &$\cdots$ \\
& 12 & {D} & $\qmark$ & $\qmark$ & {U} & {U} & {U} & {U} & {U} & {U} & {U} & {U} & {U} & {U} & {U} & {U} &$\cdots$ \\
& 13 & {D} & $\qmark$ & $\qmark$ & {U} & {U} & {U} & {U} & {U} & {U} & {U} & {U} & {U} & {U} & {U} & {U} &$\cdots$ \\ 
& 14 & {D} & $\qmark$ & $\qmark$ & {U} & {U} & {U} & {U} & {U} & {U} & {U} & {U} & {U} & {U} & {U} & {U} &$\cdots$ \\
& 15 & {D} & $\qmark$ & $\qmark$ & {U} & {U} & {U} & {U} & {U} & {U} & {U} & {U} & {U} & {U} & {U} & {U} &$\cdots$ \\
& 16 & {D} & $\qmark$ & $\qmark$ & {U} & {U} & {U} & {U} & {U} & {U} & {U} & {U} & {U} & {U} & {U} & {U} &$\cdots$ \\
& 17 & {D} & $\qmark$ & $\qmark$ & {U} & {U} & {U} & {U} & {U} & {U} & {U} & {U} & {U} & {U} & {U} & {U} &$\cdots$ \\
& 18 & {D} & $\qmark$ & {U} & {U} & {U} & {U} & {U} & {U} & {U} & {U} & {U} & {U} & {U} & {U} & {U} &$\cdots$ \\
& 19 & {D} & $\qmark$ & {U} & {U} & {U} & {U} & {U} & {U} & {U} & {U} & {U} & {U} & {U} & {U} & {U} &$\cdots$ \\
& 20 & {D} & $\qmark$ & {U} & {U} & {U} & {U} & {U} & {U} & {U} & {U} & {U} & {U} & {U} & {U} & {U} &$\cdots$ \\
& 21 & {D} & $\qmark$ & {U} & {U} & {U} & {U} & {U} & {U} & {U} & {U} & {U} & {U} & {U} & {U} & {U} &$\cdots$ \\
& 22 & {D} & $\qmark$ & {U} & {U} & {U} & {U} & {U} & {U} & {U} & {U} & {U} & {U} & {U} & {U} & {U} &$\cdots$ \\
& 23 & {D} & $\qmark$ & {U} & {U} & {U} & {U} & {U} & {U} & {U} & {U} & {U} & {U} & {U} & {U} & {U} &$\cdots$ \\
& 24 & {D} & $\qmark$ & {U} & {U} & {U} & {U} & {U} & {U} & {U} & {U} & {U} & {U} & {U} & {U} & {U} &$\cdots$ \\
& 25 & {D} & $\qmark$ & {U} & {U} & {U} & {U} & {U} & {U} & {U} & {U} & {U} & {U} & {U} & {U} & {U} &$\cdots$ \\
& 26 & {D} & $\qmark$ & {U} & {U} & {U} & {U} & {U} & {U} & {U} & {U} & {U} & {U} & {U} & {U} & {U} &$\cdots$ \\
& 27 & {D} & $\qmark$ & {U} & {U} & {U} & {U} & {U} & {U} & {U} & {U} & {U} & {U} & {U} & {U} & {U} &$\cdots$ \\
& 28 & {D} & $\qmark$ & {U} & {U} & {U} & {U} & {U} & {U} & {U} & {U} & {U} & {U} & {U} & {U} & {U} &$\cdots$ \\
& 29 & {D} & $\qmark$ & {U} & {U} & {U} & {U} & {U} & {U} & {U} & {U} & {U} & {U} & {U} & {U} & {U} &$\cdots$ \\
& 30 & {D} & $\qmark$ & {U} & {U} & {U} & {U} & {U} & {U} & {U} & {U} & {U} & {U} & {U} & {U} & {U} &$\cdots$ \\
& 31 & {D} & $\qmark$ & {U} & {U} & {U} & {U} & {U} & {U} & {U} & {U} & {U} & {U} & {U} & {U} & {U} &$\cdots$ \\
& 32 & {D} & $\qmark$ & {U} & {U} & {U} & {U} & {U} & {U} & {U} & {U} & {U} & {U} & {U} & {U} & {U} &$\cdots$ \\
& 33 & {D} & $\qmark$ & {U} & {U} & {U} & {U} & {U} & {U} & {U} & {U} & {U} & {U} & {U} & {U} & {U} &$\cdots$ \\
& 34 & {D} & $\qmark$ & {U} & {U} & {U} & {U} & {U} & {U} & {U} & {U} & {U} & {U} & {U} & {U} & {U} &$\cdots$  \\
& 35 & {D} & $\qmark$ & {U} & {U} & {U} & {U} & {U} & {U} & {U} & {U} & {U} & {U} & {U} & {U} & {U} &$\cdots$  \\
& 36 & {D} & {U} & {U} & {U} & {U} & {U} & {U} & {U} & {U} & {U} & {U} & {U} & {U} & {U} & {U} &$\cdots$ \\
& 37 & {D} & {U} & {U} & {U} & {U} & {U} & {U} & {U} & {U} & {U} & {U} & {U} & {U} & {U} & {U} &$\cdots$ \\
& 38 & {D} & {U} & {U} & {U} & {U} & {U} & {U} & {U} & {U} & {U} & {U} & {U} & {U} & {U} & {U} &$\cdots$ \\
& 39 & {D} & {U} & {U} & {U} & {U} & {U} & {U} & {U} & {U} & {U} & {U} & {U} & {U} & {U} & {U} &$\cdots$ \\
& 40 & {D} & {U} & {U} & {U} & {U} & {U} & {U} & {U} & {U} & {U} & {U} & {U} & {U} & {U} & {U} &$\cdots$ \\
&  $\vdots$ & $\vdots$ & $\vdots$ & $\vdots$ & $\vdots$ & $\vdots$ & $\vdots$ & $\vdots$ & $\vdots$ 
& $\vdots$ & $\vdots$ & $\vdots$ & $\vdots$ & $\vdots$ & $\vdots$ & $\vdots$ & $\ddots$ 
\end{tabular} }
\caption{\label{tab:free}Current knowledge about the decidability of \pbkfree{k}{\carre{\N}{d}} for all pairs $(k, d)$.}
\end{table}

It is noteworthy that Lemma~\ref{lem:inj-d-dplusun} does not hold the other way round in general.

\begin{prpstn}
Let $D$ be a semiring, 
let $K$ be a field,
and let $d$ be an integer greater than $1$.
There exists no injective morphism from $\carre{D}{d}$  to $\carre{K}{(d - 1)}$. 
\end{prpstn}

\begin{proof}
First, remark that there exists $M \in \carre{D}{d}$ such that $M^{d} \ne M^{d - 1}$ and $M^{d + 1} =  M^{d}$:
for instance, $N_d$ is a suitable choice for $M$, where $N_d$ is as in the proof of Corollary~\ref{cor:MPk}.

Now, let $n \in \N$ and $M \in \carre{K}{(d - 1)}$ be such that $M^{n + 1} = M^n$.
Let $\mu(\ttz)$ denote the minimal polynomial of $M$. 
The degree of $\mu(\ttz)$ is at most $d - 1$
and 
$\mu(\ttz)$ divides $\ttz^{n + 1} - \ttz^n = \ttz^n(\ttz - 1)$, 
so $\mu(\ttz)$ divides in fact $\ttz^{d - 1} (\ttz - 1) = \ttz^{d} - \ttz^{d - 1}$.
It follows that $M^{d} =  M^{d - 1}$.
Hence, no  $M \in  \carre{K}{(d - 1)}$ satisfies both $M^{d} \ne M^{d - 1}$ and $M^{d + 1} =  M^{d}$.
\end{proof}

To conclude the section, we put forth an interesting open question related to the decidabilities of \pbfree{\carre{\Z}{4}}
and \pbfree{\carre{\Q}{4}}.

\begin{open}[Bell and Potapov \cite{BellP08}]
Let $\Hamilton$ be as in Section~\ref{sec:to-undec-two-two}.
Is \pbfree{\Hamilton} decidable?
\end{open}

\section*{Acknowledgements}

The authors thank Juhani Karhum\"aki  for his hospitality 
and
Luc Guyot for his help in writing Section~\ref{sec:LucGuyot}.
The Academy of Finland supported the work under grants 203354 and 7523004.

\bibliography{bibmat}
\bibliographystyle{plain}

\end{document}